\documentclass{class}

\renewcommand{\sc}{\rm}

\usepackage{bm,sidecap}
\usepackage{mathrsfs,xcolor}
\usepackage{amsmath,amssymb,amsgen,amsbsy}

\newcommand{\lbreak}{\ell_{\mathrm{sc}}}
\newcommand{\Tbreak}{T_{\mathrm{sc}}}
\newcommand{\req}{r_{\rm eq}}
\newcommand{\Qeq}{Q_{\rm eq}}
\newcommand{\etaK}{\eta_{\rm K}}

\newcommand{\Wi}{\mathit{Wi}}
\newcommand{\Wic}{\mathit{Wi}_{\rm cr}}

\title{Polymer scission in turbulent flows}

\author{
Dario Vincenzi\aff{1},
Takeshi Watanabe\aff{2},  
Samriddhi Sankar Ray\aff{3},
\and 
Jason~R.~Picardo\aff{4}
}

\affiliation{
\aff{1}Universit\'e C\^ote d'Azur, CNRS, LJAD, 06100 Nice, France
\aff{2}Department of Physical Science and Engineering, Nagoya Institute of Technology, Gokiso, Nagoya 466-8555, Japan
\aff{3}International Centre for Theoretical Sciences, 
Tata Institute of Fundamental Research, Bangalore 560089, India
\aff{4} Department of Chemical Engineering, Indian Institute of Technology Bombay, Mumbai 400076, India
}

\begin{document}

\maketitle

\begin{abstract}
Polymers in a turbulent flow are subject to intense strain, which can cause their scission
and thereby limit the experimental study and application of phenomena such as turbulent
drag reduction and elastic turbulence. In this paper, we study polymer scission in
homogeneous isotropic turbulence, through a combination of stochastic modelling,
based
on a Gaussian time-decorrelated random flow, and direct numerical simulations (DNSs)
with both one-way (passive) and two-way (active) coupling of the polymers and the flow. For the first scission of
passive polymers, the stochastic model yields analytical predictions which are
found to be in good agreement with results from the DNSs, for the
temporal evolution of the fraction of unbroken polymers and the statistics of the survival
of polymers. The impact of scission on the dynamics of a turbulent polymer solution is investigated through DNSs with
two-way coupling (active polymers). Our results
indicate that the reduction of kinetic energy dissipation due to feedback from stretched polymers is an inherently transient effect, which is lost as the polymers breakup. Thus, the overall dissipation-reduction is maximised by an intermediate polymer relaxation time, for which polymers stretch significantly but without breaking too quickly. We also study the dynamics of the polymer fragments which form after scission; these daughter polymers can themselves undergo subsequent, repeated, breakups to produce a hierarchical population of polymers with a range of relaxation times and scission rates. 
\end{abstract}

\section{Introduction}

The viscoelastic properties of dilute polymer solutions are central to
several physical and engineering applications~\citep{l99}.
When the pure solvent is turbulent the most remarkable effect of
the addition of polymers is a significant
reduction of the turbulent drag below that of
the solvent \citep*{plb08,wm08,b10,g14}. 
This phenomenon, also 
known as Toms effect \citep{Toms1,Toms2},
is commonly utilized to reduce the energy losses in pipelines
and hence the costs associated with the transport of crude oil.
However, in a turbulent flow,
polymers are subject to mechanical degradation due to 
the fluctuating strain-rate, which stretches polymers and thus causes 
their scission.
Since turbulent drag reduction decreases with the molecular weight of the
dissolved polymers \citep{v75}, 
the efficacy of polymers as drag reducing agents
diminishes in time, with
a strong impact on both industrial applications and laboratory experiments
\citep*{PA70,dTDKN95,cllc02,vis05,vcs06,ewsc09,ps12,odp17}.

An analogous mechanical
degradation of polymers and, by association, practical limitations
are observed in experiments of
homogeneous and isotropic turbulence with polymer additives \citep{cmxb08}
as well as in the regime of elastic turbulence \citep{gs04}. In the
latter case, even though the Reynolds number of the solution is low, elastic
instabilities generate a chaotic flow with highly fluctuating
velocity gradients \citep*{bss07}.

A detailed knowledge of the statistics of
polymer scission in turbulent flows
is therefore important for the design of experiments
and the performance of realistic simulations of both
turbulent drag reduction
and elastic turbulence.
However, unlike the fragmentation of liquid sheets and drops \citep{villermaux},
the modelling of flow-driven scission of polymers has received scant attention and, with very few exceptions \citep*[e.g.][]{pms18},
polymer scission is disregarded completely by the available constitutive models of polymer solutions.

We therefore investigate the dynamics of polymers in
a turbulent flow focusing on the scission statistics.
A polymer is described as a bead-spring chain in a time-dependent,
linear velocity field.
This polymer model is known as the \citet{r53} model
and represents one of the most common descriptions
of a polymer molecule in a flow
(in the case in which
only two beads are considered, the Rouse model
reduces to the elastic dumbbell model; see  \citealt{bcah77}).
Even when the flow is turbulent,
the assumption of a linear velocity field is justified,
since the size of polymers is generally smaller than the
Kolmogorov dissipation scale $\eta_K$, below which viscosity strongly damps
the spatial fluctuations of the velocity.
We introduce scission into the Rouse model by assuming that 
the bead-spring chain breaks into two shorter chains
as soon as the tension in one of the springs
exceeds a critical threshold.

After introducing a modification of the Rouse model that takes polymer scission into account,
we begin by considering the statistics of the first scission.
For {\it passively} transported polymers,
for which the motion of the polymers does not modify the carrier velocity field,
we derive qualitative analytical predictions
by restricting ourselves to the Hookean dumbbell model and by
using a decorrelated-in-time
Gaussian stochastic velocity field (the approach is adapted from
a study of droplet breakup conducted in \citealt{RV18}).
The theoretical predictions are compared with Lagrangian direct numerical
simulations (DNSs) of the Rouse model in three-dimensional homogeneous isotropic turbulence.
For {\em active} polymers, the statistics of the first scission is studied via
hybrid Eulerian--Lagrangian simulations \citep{wg13jfm,wg13b,wg14},
in which the feedback of (dumbbell-like) polymers on the velocity field is taken into account. 
These simulations shed light on the transient nature of the dissipation-reduction effect, which owes its origin to polymer stretching and its demise to polymer scission. 
Finally, we analyse multiple scissions of passive polymers, via DNSs in which a hierarchy of daughter polymers arise from successive breakups, each of
which with their own statistics.

\section{The Rouse chain}

The Rouse model describes a polymer as a chain of $\mathscr{N}$ inertialess beads connected
to their nearest neighbors by elastic springs.
We consider FENE (finitely extensible nonlinear elastic)
springs with spring constant $H$ and maximum length $Q_m$.
The fluid in which the chain is immersed is Newtonian and its motion is described
by an incompressible velocity field $\bm u(\bm x,t)$.
The drag force of the fluid on each bead is given by the Stokes law
with drag coefficient $\zeta$; the collisions of the molecules of the fluid
with a bead are described by Brownian motion.
Finally, hydrodynamic and excluded-volume interactions between different segments of
the chain are disregarded.

The motion of the chain is  described
in terms of the position of its center of mass, $\bm X_c$, and the separation
vectors between the beads, $\bm Q_i$ ($i=1,\dots,\mathscr{N}-1$).
This set of coordinates evolves according to the equations \citep{bcah77,o96}%
\begin{subequations}%
\begin{eqnarray}%
  \label{eq:cm}
  \dot{\bm X}_c &=& \bm u(\bm X_c(t),t)+\dfrac{1}{\mathscr{N}}\sqrt{\frac{\Qeq^2}{6\tau}}
  \sum_{i=1}^{\mathscr{N}}\bm\xi_i(t),
  \\[2mm]
  \label{eq:q}
  \dot{\bm Q_i} &=& 
  \bm\kappa(t)\bcdot\bm Q_i(t)-\dfrac{1}{4\tau}
           [2f_i\bm Q_i(t)-f_{i+1}\bm Q_{i+1}(t)-f_{i-1}\bm Q_{i-1}(t)]
           \\
           \nonumber
           &&+\sqrt{\dfrac{\Qeq^2}{6\tau}}
             [\bm\xi_{i+1}(t)-\bm\xi_i(t)],\qquad i=1,\dots,\mathscr{N}-1,%
\end{eqnarray}%
\label{eq:chain}%
\end{subequations}%
where $\kappa^{\alpha\beta}(t)=\nabla^\beta u^\alpha(\bm X_c(t),t)$
is the velocity gradient evaluated at the position of the center of mass,
$\tau=\zeta/4H$
is the characteristic time scale of the springs, $\Qeq=\sqrt{3k_BT/H}$ is their
equilibrium root-mean-square (r.m.s.)
extension ($k_B$ denotes the Boltzmann constant
and $T$ is temperature),
and $\bm\xi_i(t)$
($i=1,\dots,\mathscr{N}$) are independent, vectorial, white noises. 
The coefficients
\begin{equation}
f_i=\dfrac{1}{1-\vert\bm Q_i\vert^2/Q_m^2}
\end{equation}
characterize the FENE interactions and
ensure that the extension of each spring 
does not exceed its maximum length $Q_m$.
Obviously,
in the equations for $\bm Q_1$ and $\bm Q_{\mathscr{N}-1}$ it is assumed that 
$\bm Q_0=\bm Q_{\mathscr{N}}=0$. 

The end-to-end separation vector of the polymer is
defined as $\bm R=\sum_{i=1}^{\mathscr{N}-1}\bm Q_i$.
In a still fluid, the equilibrim r.m.s. value of $\vert\bm R\vert$ is
$\req=\Qeq\sqrt{\mathscr{N}-1}$ \citep{bcah77}.

We modify the Rouse model in order to account for the scission of the polymer
when the tension in any of the springs exceeds a critical value.
Since the relation between the tension and the extension of a spring
can be easily inverted~\citep{thiffeault}, we can, equivalently, assume
that for each spring of the chain there exists a critical scission length
$\lbreak$ such that the spring breaks
if the length of the corresponding separation vector
exceeds $\lbreak$ (i.e. the chain breaks if
$\vert\bm Q_i\vert\geqslant\lbreak$ for any $1\leqslant i\leqslant \mathscr{N}-1$).

The scission process is non-stationary; the dynamics of the chain
therefore depends on its initial configuration and, in particular, on
its initial end-to-end separation $r_0=\vert\bm R(0)\vert$.
In the following we shall assume
that $\lbreak$ is much greater than $r_0/(\mathscr{N}-1)$ and that
$r_0$ is equal to $\req$
or greater than it.

Finally, the size of the chain always remains
smaller than $\etaK$, so that the velocity field at the scale of the
chain can be considered as linear and the dynamics of the polymer is
entirely determined by the velocity gradient at the location of the
center of mass, consistent with the Rouse model.

To summarize, the spatial scales that characterize 
the system are arranged as follows:
$\req\leqslant r_0\ll\lbreak(\mathscr{N}-1)<Q_m(\mathscr{N}-1)<\etaK$. 

\section{First-scission statistics}

\subsection{Passive polymers}
\label{sect:first-scission}

\subsubsection{Analytical predictions}
\label{sect:analytical}

Here we make some simplifying assumptions on both the polymer model and the
carrier flow in
order to derive analytical predictions for the statistics of polymer scission. 

First of all, we only consider the statistics of the first scission.
We then restrict ourselves 
to the $\mathscr{N}=2$ case, also known as the dumbbell
model \citep{bcah77,o96},
i.e., we focus on the slowest deformation mode of the polymer.
Many results on single-polymer dynamics in random or turbulent flows
have been obtained by using the
dumbbell model (see \citealt{VPBT15} and references therein) 
and the most common constitutive models of polymer
solutions, namely the Oldroyd-B \citep{o50} and the FENE-P \citep{bdj80} models,  are based on it.
The legitimacy of this approach is supported by the numerical simulations
in \cite{jc08} and~\cite{wg10}, where it is shown that, in isotropic turbulence
and in the absence of scission,
the statistics of the end-to-end separation of a dumbbell and that of a $\mathscr{N}=20$ chain
coincide (provided, of course,
a proper mapping between the parameters of the two systems is applied).
Finally, we replace the nonlinear spring with a Hookean one ($f_i=1$); 
this is because the nonlinearity
of the elastic force enters into play only at extensions 
close to the scission length, 
and we shall see from our simulations in \S~\ref{sect:sim}
that it does not affect
the qualitative properties of the scission process.
For $\mathscr{N}=2$, \eqref{eq:chain} reduce to
\begin{subequations}
\begin{eqnarray}
  \label{eq:cm-dumbbell}
  \dot{\bm X}_c
  &=& \bm u(\bm X_c(t),t)+\dfrac{1}{2}\sqrt{\frac{\req^2}{3\tau}}
  \,\bm \zeta_1(t),
  \\
  \label{eq:q-dumbbell}
  \dot{\bm R} &=& \bm\kappa(t)\bcdot\bm R(t)-\dfrac{\bm R(t)}{2\tau}
      +\sqrt{\dfrac{\req^2}{3\tau}}\,
      \bm\zeta_2(t),
\end{eqnarray}
\label{eq:dumbbell}
\end{subequations}
where $\bm\zeta_1(t)$ and $\bm\zeta_2(t)$ are (non-independent)
vectorial white noises.

We model the flow via the smooth (also known as Batchelor) regime
of the \cite{k68} model.
This model
has been widely employed in the study of turbulent transport \citep*{fgv01}
and has yielded several theoretical results
on the coil-stretch transition in random or turbulent flows
(see \citealt*{pav16} and references therein). 
The velocity is a divergence-less and spatially smooth
Gaussian vector field.
It is statistically stationary in time and
homogeneous, isotropic, and parity invariant in space;
it has zero mean and zero correlation time.
Under these assumptions, 
$\bm \kappa(t)$ is a tensorial white noise with two-time correlation
\citep{fgv01}
\begin{equation}
  \langle \kappa^{ij}(t)\kappa^{mn}(t')\rangle=
  \mathscr{K}^{ijmn}\delta(t-t'),
\end{equation}
where
\begin{equation}
  \mathscr{K}^{ijmn}=\frac{\lambda}{3}
          [4\delta^{im}\delta^{jn}-\delta^{ij}
            \delta^{mn}-\delta^{in}\delta^{jm}]
\end{equation}
and $\lambda$ is the maximum Lyapunov exponent of the flow.
Obviously, the assumption of temporal decorrelation is a strong
approximation, since an isotropic turbulent flow has Kubo number 
$\mathit{Ku}=\lambda t_{\mathrm{corr}}\approx 0.6$,
where $t_{\mathrm{corr}}$ is the correlation time of the flow \citep{gp90,bec06,wg10}.
However, it was shown in \cite{mv11} that, for a
fluctuating flow with comparable $\mathit{Ku}$,
the statistics of polymer extension
is correctly captured by a time decorrelated velocity field.

The Weissenberg number $\Wi=\lambda\tau$ determines 
to what extent polymers are stretched by the flow. In particular,
the coil-stretch transition occurs when $\Wi$ exceeds the critical value $\Wic=1/2$ 
(\citealt*{lumley,bfl00}; note that our definition of $\tau$ 
and hence that of $\Wi$ differ from that of \citealt{bfl00} by a factor of~2).

As the velocity field is homogeneous and isotropic in space, the statistics of
$R=\vert\bm R\vert$
is independent of the position of the center of mass and of the
direction of $\bm R$.
To study polymer scission,
it is therefore sufficient to focus on 
the probability density function (p.d.f.) of $R$, which will be denoted as $P(R,t)$.
When $\bm\kappa(t)$ has the properties described above,
$P(R,t)$ satisfies the Fokker--Planck equation \citep*{c00,cmv05}:
\begin{equation}
\label{eq:FPE}
\partial_{t'} P=\mathbb{L} P, 
\qquad \mathbb{L} P=- \partial_R (D_1 P)+\partial_R^2(D_2 P),
\end{equation}
with rescaled time $t'=t/2\tau$ and coefficients
\begin{equation}
  \label{eq:coeff}
  D_1 = \Big(\frac{8}{3}\Wi-1\Big)R+\dfrac{2\req^2}{3R}, \qquad
  D_2 = \frac{2\Wi}{3}R^2+\frac{\req^2}{3}
\end{equation}
(once again our definition of $\Wi$
differs from that used in \citealt{c00} and \citealt{cmv05}
by a factor of 2).
The appropriate boundary conditions are
reflecting at $R=0$ and absorbing at $R=\lbreak$, i.e. 
\begin{equation}
  \label{eq:bc}
  D_1 P-\partial_R(D_2 P)=0 \ \text{at $R=0$}
  \qquad \text{and} \qquad
  P(\lbreak,t)=0
  \end{equation}
for all $t$.
The former condition ensures that the extension of the polymer
stays positive, while the latter describes scission at $R=\lbreak$.
The analysis of \eqref{eq:FPE} to~\eqref{eq:bc}
closely follows that in \citet{RV18}
for the breakup of sub-Kolmogorov droplets in isotropic turbulence.
(The results presented here are deduced directly from those in \S~3 of
\citealt{RV18} by setting $\mathit{Ca}=\Wi$, $\mu=1$, $r_{\mathrm{eq}}^2=
  \Qeq^2/3$, and
  $f_1(\mu)=f_2(\mu)=\gamma(\mu)=1$.)
We therefore skip the details of the derivations and 
directly present the predictions of scission statistics.

The number of unbroken polymers that survive at time $t$,
$N_p(t)$,
is related to $P(R,t)$ as follows:
\begin{equation}
N_p(t)/N_p(0)=\int_0^{\lbreak} P(R,t)\mathrm{d}R.
\label{eq:N}
\end{equation}
At times $t\gg \tau$,
$N_p(t)$ therefore decays exponentially as
\begin{equation}
N_p(t)\sim N_p(0)\,e^{-t/T_d},
\label{eq:N-decay}
\end{equation}
where the decay time $T_d$ is the reciprocal of the
lowest eigenvalue of the operator $\mathbb{L}$ with boundary conditions
\eqref{eq:bc}.
The eigenfunctions of $\mathbb{L}$ are hypergeometric functions with
parameters depending on $\req$, $r_0$, $\Wi$ and form a discrete set
selected by the boundary condition at $R=\lbreak$.
A calculation of the lowest eigenvalue shows that
$T_d$ depends weakly on $\Wi$ for small $\Wi$,
decreases rapidly as $\Wi$ exceeds $\Wic$, and saturates at large $\Wi$. 

As we shall see below, the p.d.f. of $R$ integrated over time,
\begin{equation}
\label{eq:Phat}
\widehat{P}(R)=\int_0^\infty P(R,t)\mathrm{d}t,
\end{equation}
allows us to estimate the mean lifetime of a polymer before its first 
scission. 
If the initial distribution of polymer sizes is `monodisperse',
i.e. $P(R,0)=\delta(R-r_0)$, then $\widehat{P}(R)$ takes the form
\begin{equation}
  \widehat{P}(R)\propto
  \begin{cases}
    e^{-\Phi(R)}[\phi(\lbreak)-\phi(r_0)] & \text{if $0\leqslant R\leqslant r_0$},
    \\
    e^{-\Phi(R)}[\phi(\lbreak)-\phi(R)] & \text{if $r_0< R\leqslant\lbreak$}
  \end{cases}
\end{equation}
with
\begin{equation}
  \Phi(R)=\ln D_2(R)-\int^R \dfrac{D_1(\zeta)}{D_2(\zeta)}\,\mathrm{d}\zeta,
  \qquad
  \phi(R)=\int^R \dfrac{e^\Phi(\zeta)}{D_2(\zeta)}\,\mathrm{d}\zeta
\end{equation}
and hence
\begin{equation}
\label{eq:pdf-1}
\widehat{P}(R)\sim
\begin{cases}
\req^{-2}r_0^ {-1}R^2 & \text{if $0\leqslant R\ll \req$,} 
\\[1mm]
\vert r_0^\alpha-\lbreak^\alpha\vert
R^{-1-\alpha} & \text{if $\req\ll R\ll r_0$,} 
\\[1mm]
\lbreak^\beta R^{-1-\beta} & \text{if $r_0\ll R\ll\lbreak$},
\end{cases}
\end{equation} 
where $\alpha=3(\Wi^{-1}-2)/2$ and
\begin{equation}
\beta = \begin{cases}
\alpha & \text{if $\Wi<\Wic$,}
\\
0 & \text{if $\Wi>\Wic$}.
\end{cases}
\end{equation}
Therefore, above the coil-stretch transition,
the right tail of $\widehat{P}(R)$ saturates to the power law $R^{-1}$.
An analogous behaviour was found previously for the size distribution of
sub-Kolmogorov droplets in isotropic turbulence \citep*{bmv14,RV18}.
Also note that the exponent 
$\alpha$ coincides with the one obtained by \cite{bfl00}
for the p.d.f. of intermediate extensions
in the absence of scission.

If the initial distribution of polymer sizes is broad but nonetheless
admits a maximum size $r_0$, then the left ($R\ll\req$)
and right ($r_0\ll R\ll\lbreak$) power-law tails continue to exist, but
$\widehat{P}(R)$ no longer behaves as a power law
for extensions $\req\ll R\ll r_0$.

The mean time $\langle\Tbreak\rangle$ it takes for a polymer
to undergo its first scission can be deduced from the behaviour of
$\widehat{P}(R)$ via the relation:
\begin{equation}
\langle\Tbreak\rangle=\int_0^{\lbreak}\widehat{P}(R)\mathrm{d}R.
\end{equation}
Equation \eqref{eq:pdf-1} then yields
two different behaviours below and above the coil-stretch transition:
\begin{equation}
  \label{eq:tsc}
\lambda\langle\Tbreak\rangle\sim
\begin{cases}
(\lbreak/r_0)^\beta & \text{if $\Wi<\Wi_{\rm cr}$},
\\
\ln (\lbreak/r_0) & \text{if $\Wi>\Wi_{\rm cr}$}.
\end{cases}
\end{equation}

\subsubsection{Direct numerical simulations}
\label{sect:sim}

In this Section,
we present numerical simulations of the Rouse model \eqref{eq:chain} 
in homogeneous and isotropic turbulence and compare them with 
the analytical predictions of \S~\ref{sect:analytical}.
The velocity field $\bm u(\bm x,t)$
is the solution of the incompressible Navier--Stokes equations,
\begin{equation}
\partial_t \bm u+\bm u\bcdot\bnabla\bm u=-\bnabla p+\nu\Delta\bm u+\bm F,
\quad \bnabla\bcdot\bm u=0,
\label{eq:NS}
\end{equation}
over the periodic cube $[0,2\upi]^3$. Here $p$ is pressure, $\nu$ is the
kinematic viscosity, and $\bm F(\bm x,t)$ a body force that mantains a constant
kinetic-energy input $\epsilon$. 
The numerical integration uses a standard, fully de-aliased pseudo-spectral
method with $512^3$ collocation points and, for the time evolution,
 a second-order slaved  Adams--Bashforth scheme 
with time step ${\rm d}t=4\times 10^{-4}$.
The values of $\nu$ and $\epsilon$ are such
that the Taylor-microscale Reynolds number is $\mathit{Re}_\lambda=111$.
The Lyapunov exponent of the flow is $\lambda=0.15\tau_\eta^{-1}$, where
$\tau_\eta$ denotes the Kolmogorov time scale, consistent with the 
value found earlier in \citet{bec06} and \citet{wg10}.

The position of the centre of mass of the polymer is obtained by 
integrating~\eqref{eq:cm} via a second-order Adam--Bashforth method with same
${\rm d}t$ as for the Navier--Stokes equations. 
The noise term in~\eqref{eq:cm}
is disregarded, because it has a negligible effect when $\bm u(\bm x,t)$
is turbulent. Moreover, its amplitude 
is smaller than that of the noise terms in the
equations for the separation vectors by a factor of $\mathscr{N}$. 
As $\bm u(\bm x,t)$ is only known
over a discrete grid, the integration of~\eqref{eq:cm} requires 
interpolation to reconstruct the velocity field at 
$\bm X_c(t)$; a trilinear scheme is used for this purpose.
The same approach allows the calculation of the velocity gradient
along the trajectory of the center of mass, $\bm \kappa(t)$, and hence
the integration of~\eqref{eq:q} 
by means of the Euler--Maruyama method with time step ${\rm d}t$.
Since we focus on the statistics of the first scission, a chain is removed
from the simulation as soon as it breaks according to the criterion
discussed above.
The time origin for \eqref{eq:q} ($t=0$) is taken in the statistically steady state
of the carrier turbulent flow, so that the temporal dynamics of polymers
is not influenced by the initial transient evolution of $\bm u(\bm x,t)$.
Note that, in the present context, it is not necessary to use
integration schemes specifically designed to
prevent the extension of the links from exceeding $Q_m$, 
since the springs, by construction, break well before their extension approaches $Q_m$.

\begin{figure}
\includegraphics[width=0.49\textwidth]{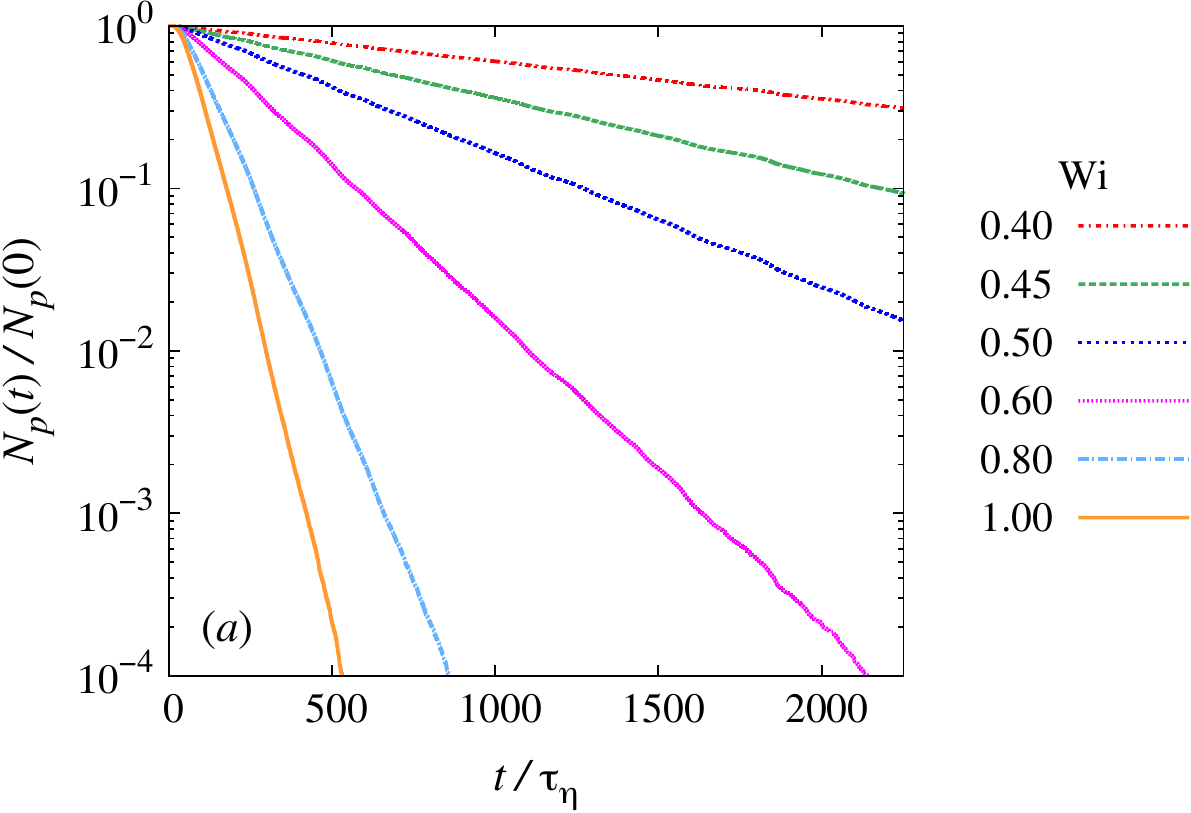}\hfill%
\includegraphics[width=0.49\textwidth]{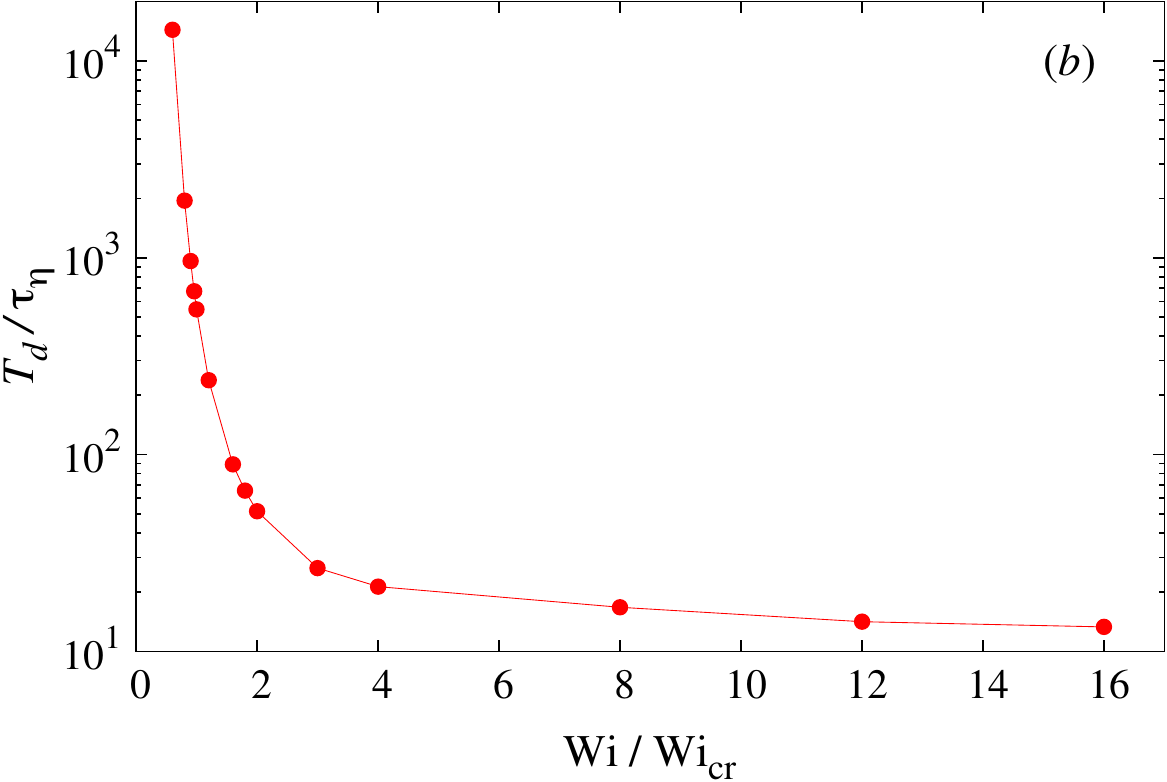}%
\caption{Passive polymers:
(\textit{a}) exponential decay of the fraction of unbroken polymers for different values of $\Wi$;
(\textit{b}) decay time of the fraction of unbroken polymers rescaled by the
Kolmogorov time $\tau_\eta$ as a function
of $\Wi/\Wi_{\rm cr}$.}
\label{fig:decay}
\end{figure}

In our simulations, we consider
$N_p(0)=9\times 10^5$ polymers, whose positions at time $t=0$ are
uniformly distributed in space.
Since the statistics of polymer extension depends on the initial
size of polymers but not on their orientation, for simplicity the 
initial condition for the separation vectors is taken to be
$\bm Q_i(0)=Q_0 (1,1,1)/\sqrt{3}$ with $Q_0>0$ for all polymers, i.e.
the polymers
are in a straight configuration and 
$P(R,0)=\delta(R-r_0)$ with $r_0=(\mathscr{N}-1)Q_0$.

In order to compare chains with different numbers of beads,
an appropriate mapping of the chain parameters is needed.
Here we use the mapping proposed by \cite{jc08} and also used
by \cite{wg10}. 
If the parameters of the individual links of a $\mathscr{N}$-bead chain are
$\tau$, $\Qeq$, $Q_m$, $\lbreak$, then the statistics of the end-to-end
separation of the chain is equivalent to that of a dumbbell with 
following parameters:
\begin{equation}
\label{eq:mapping}
\tau^{D}=\frac{\mathscr{N}(\mathscr{N}+1)\tau}{6}, \quad 
\Qeq^D=\Qeq, \quad
Q_m^{D}=Q_m{\sqrt{\mathscr{N}-1}}, \quad
\lbreak^{D}=\lbreak{\sqrt{\mathscr{N}-1}},
\end{equation}
where the last relation is introduced for 
compatibility with the expression of $Q_m^D$.
This mapping allows us to compare chains with different numbers of beads
by using the dumbbell model as a reference.

Following \cite{wg10}, we 
define the Weissenberg number for a Rouse chain
as $\Wi=\lambda\tau^{D}$. 
In our simulations, $0.4\leqslant \Wi
\leqslant 8$.
(Note that small-$\Wi$ simulations are computationally more demanding, because
the calculation of quantities like $\widehat{P}(R)$ and 
$\langle\Tbreak\rangle$ requires that the time evolution is
long enough for all polymers to break, and the time at which the scission
process is complete becomes longer
and longer as $\Wi$ decreases.)
%
\begin{figure*}
\setlength{\unitlength}{\textwidth}
\includegraphics[width=0.49\textwidth]{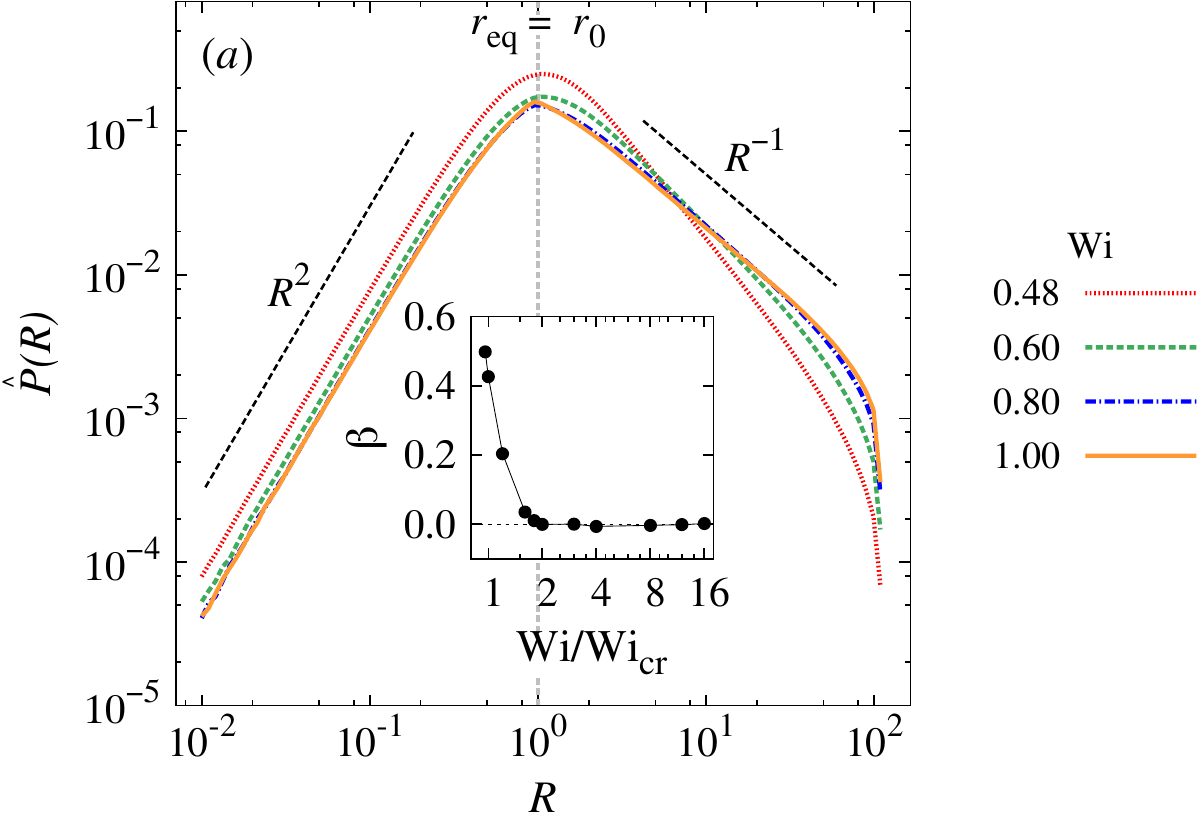}\hfill%
\includegraphics[width=0.49\textwidth]{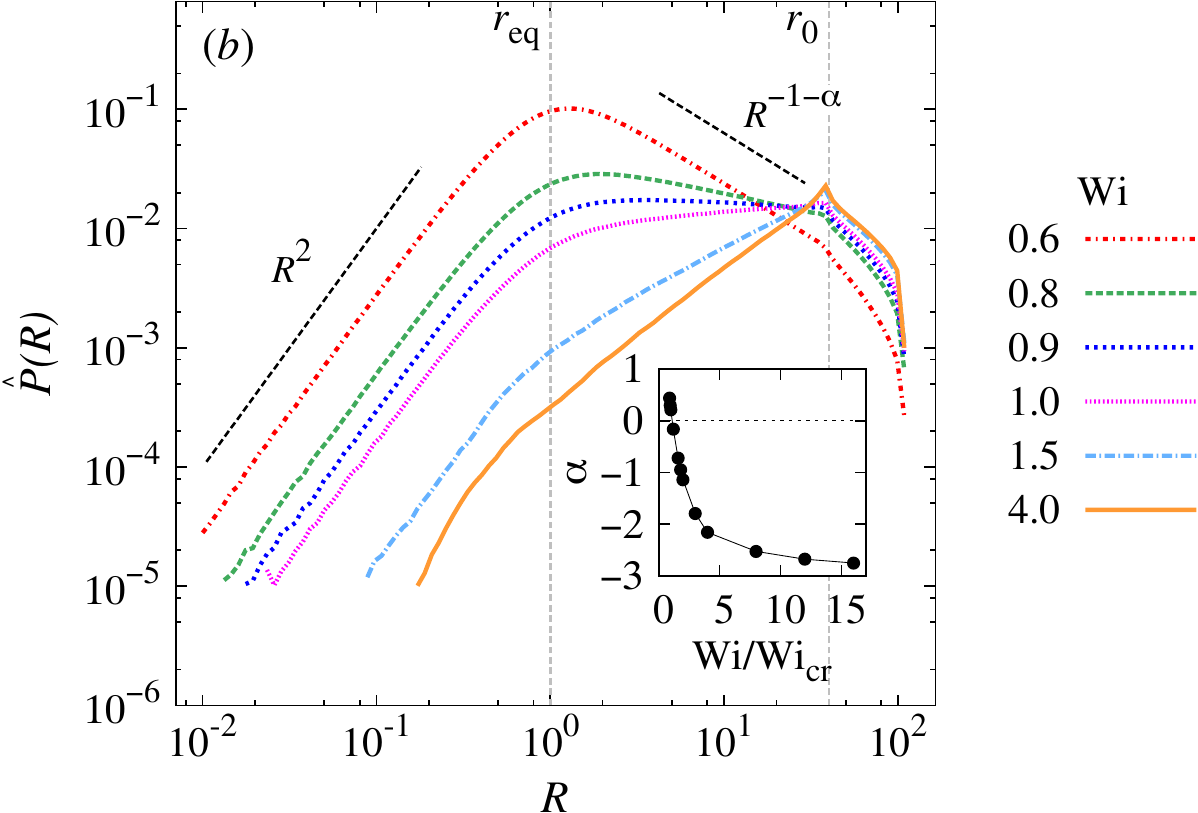}%
\caption{Passive polymers:
(\textit{a}) time-integrated p.d.f. of the end-to-end separation of unbroken polymers
for $r_0=\req$ and for different values of $\Wi$. The inset shows the
value of $\beta$, which determines the exponent of the right tail
of the p.d.f. (i.e. $\widehat{P}(R)\propto R^{-1-\beta}$
for $r_0\ll R\ll\lbreak$),
as a function of $\Wi/\Wi_{\rm cr}$;
(\textit{b}) time-integrated p.d.f. of the end-to-end separation of unbroken polymers
for $r_0=40\req$ and for different values of $\Wi$.
The inset shows the
value of $\alpha$,
which determines the power-law behaviour
of the p.d.f. for intermediate extensions (i.e.
$\widehat{P}(R)\propto R^{-1-\alpha}$ for
$\req\ll R\ll r_0$),
as a function of $\Wi/\Wi_{\rm cr}$.
In both panels (\textit{a}) and (\textit{b}), the p.d.f.s are
normalized to unity for the sake of comparison.
}
\label{fig:pdfs}
\end{figure*}

As for the choice of the other parameters, 
we take $\req=1$, $Q_m^{D}=\sqrt{3000}$
(also following \citealt{jc08}  and \citealt{wg10}), and, unless otherwise
specified, $r_0=\req$. In addition,
it is assumed that a spring breaks as soon as 
its extension exceeds $\lbreak=0.8Q_m$.
The number of beads is set to $\mathscr{N}=10$.
We have also performed simulations with different sets of parameters,
which support the generality of the results presented below.

Figure~\ref{fig:decay} shows the temporal evolution of
the fraction of unbroken polymers.
The decay is exponential 
with a time scale $T_d$ that decreases rapidly as $\Wi$ 
exceeds its critical value, in agreement with the predictions of
\S~\ref{sect:analytical}. We shall see in \S~\ref{sect:active}
that, for a dumbbell, it is possible to write an explicit expression
for $T_d$ as a function of $\Wi$.

\begin{figure*}
\centering
\includegraphics[width=0.483\textwidth]{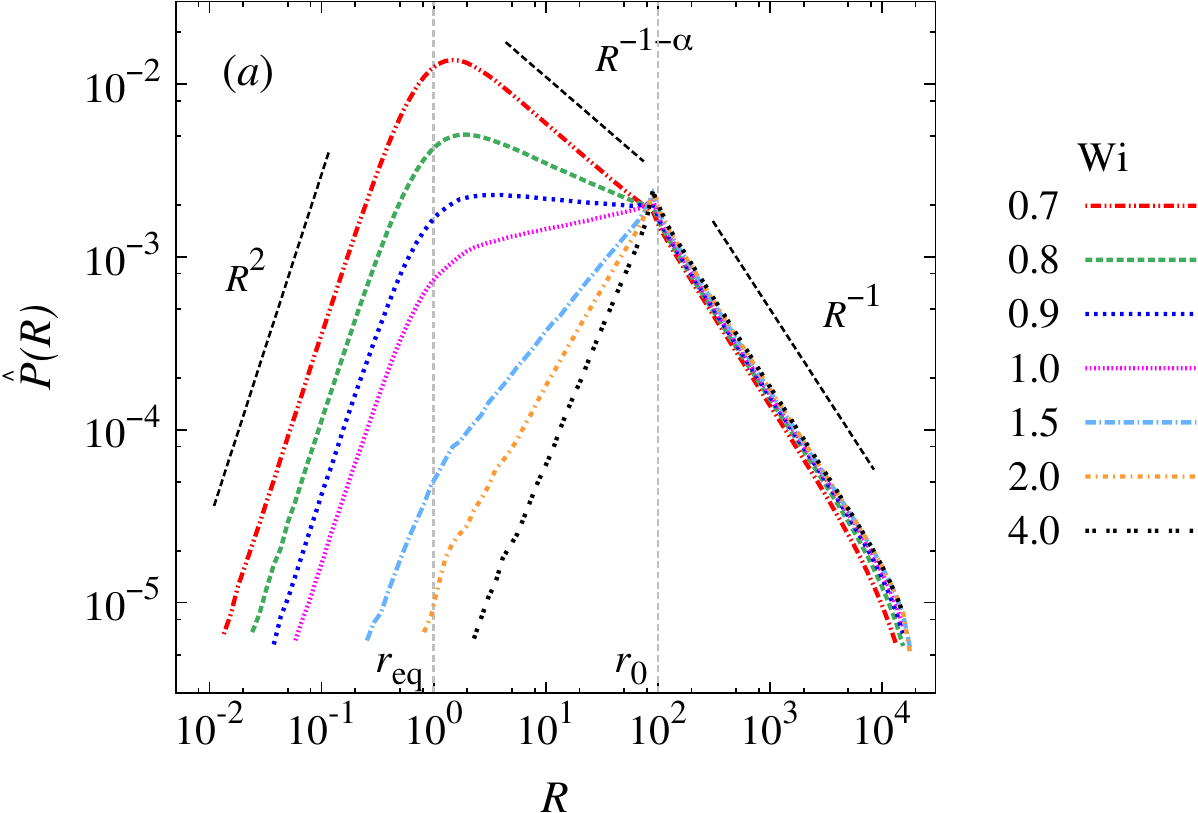}\hfill
\includegraphics[width=0.5\textwidth]{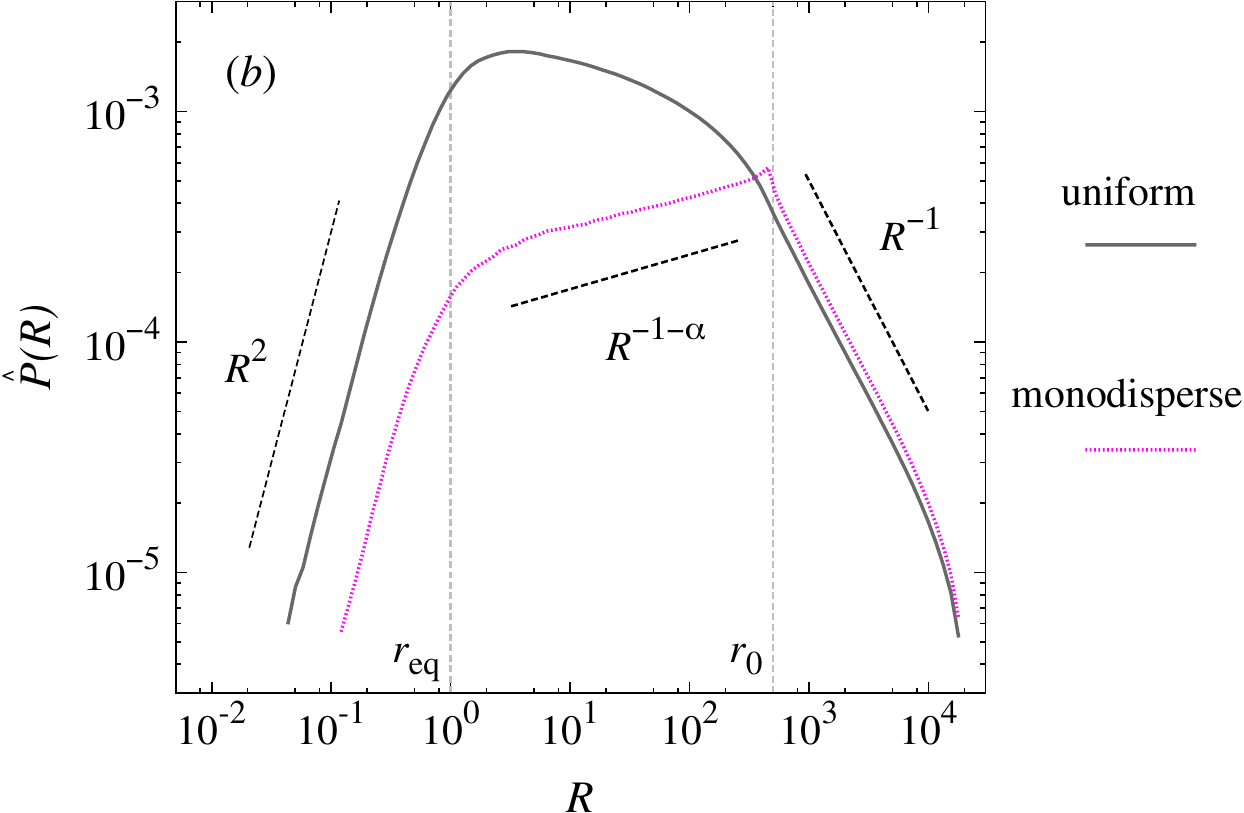}%
\caption{Passive polymers:
(\textit{a}) time-integrated p.d.f. of the end-to-end separation of unbroken polymers
  for $Q^D_m=10^4$, $\req=1$, $r_0=10^2$, and different values of $\Wi$;
  (\textit{b}) time-integrated p.d.f. of the end-to-end separation of unbroken 
polymers
for $Q^{D}_m=10^4$, $\req=1$, $r_0=5\times 10^2$, $\Wi=1$, and
a uniform (solid gray line) or a monodisperse (dashed magenta line)
initial distribution of polymer sizes. In both panels,
the p.d.f.s are
normalized to unity for the sake of comparison.}
\label{fig:ld_1e4}
\end{figure*}

The time-integrated p.d.f. of the end-to-end separation 
{\it of unbroken polymers}
is shown in figure~\ref{fig:pdfs}\textit{a}
for an initial polymer size $r_0=\req$ and different values of $\Wi$.
The p.d.f. displays a power-law behaviour
for both $R\ll \req$ and $r_0=\req\ll R\ll \lbreak(\mathscr{N}-1)$. The left tail is
proportional to $R^2$, because
the small separations are dominated by thermal fluctuations.
The right tail rises as a function
of $\Wi$, until the power law saturates to $R^{-1}$ for $\Wi>\Wi_{\rm cr}$. 
A third power law emerges for intermediate extensions
if $r_0>\req$ (figure~\ref{fig:pdfs}\textit{b}).
In this case, the exponent $-1-\alpha$ changes from negative to positive
as $\Wi$ increases and saturates to $2$ at large $\Wi$.
To appreciate the coexistence of these three power laws more clearly,
in figure~\ref{fig:ld_1e4}\textit{a} we also consider $\widehat{P}(R)$ for 
a much larger value of $Q_m^D$ and
a larger separation between $\req$, $r_0$, and $(\mathscr{N}-1)\lbreak$.
All these results confirm the predictions reported in \S~\ref{sect:analytical}.

The p.d.f.s presented so far correspond to a ``monodisperse'' initial
state $P(R)=\delta(R-r_0)$ in which all polymers have the same
end-to-end distance. However,
as mentioned in \S~\ref{sect:analytical}, the behaviour of
$\widehat{P}(R)$ for intermediate extensions
is expected to change if the initial distribution of polymer
extensions is broad. To confirm this prediction, we have considered an initial
state in which the end-to-end distance of polymers is
distributed uniformly between $\req$ and a maximum initial extension
$r_0>\req$. The time-integrated
p.d.f.s given in figure~\ref{fig:ld_1e4}\textit{b} show that
only the left and right power-law tails persist in this case,
while $\widehat{P}(R)$
does not behave as a power law for intermediate extensions.
%

\begin{figure}
\includegraphics[width=0.48\textwidth]{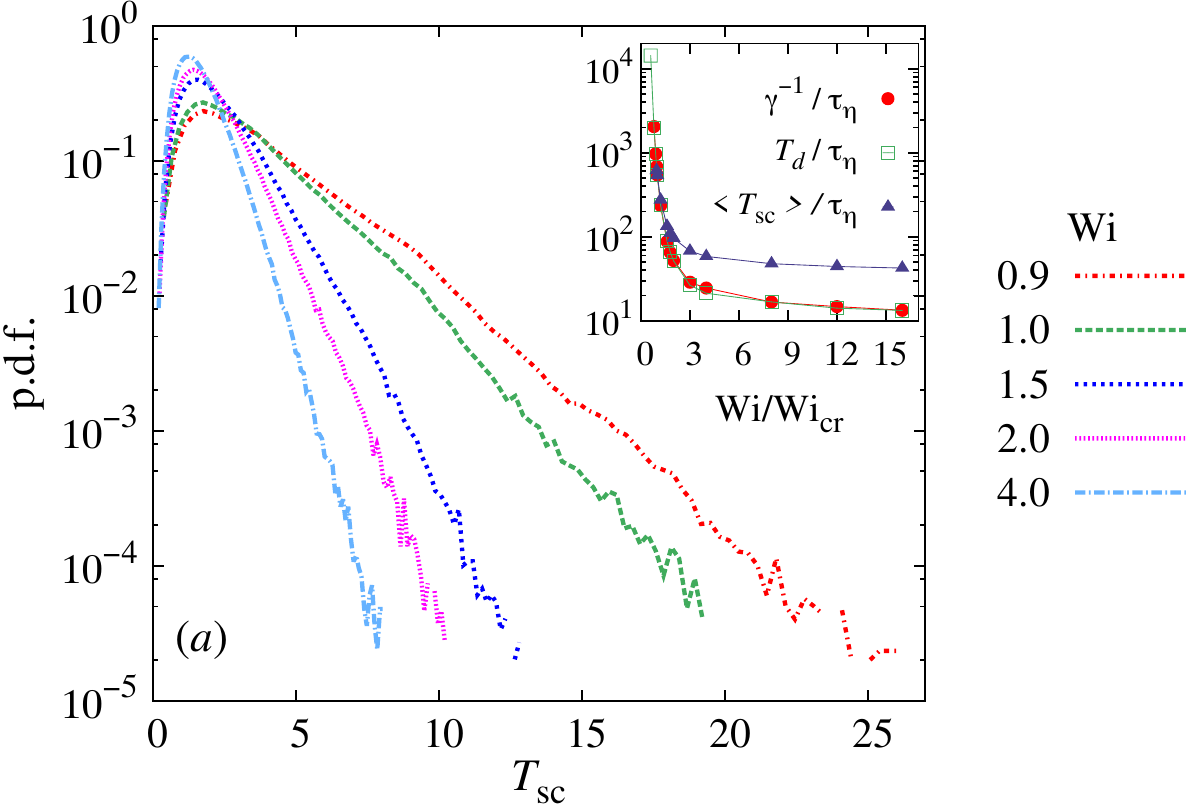}\hfill%
\includegraphics[width=0.48\textwidth]{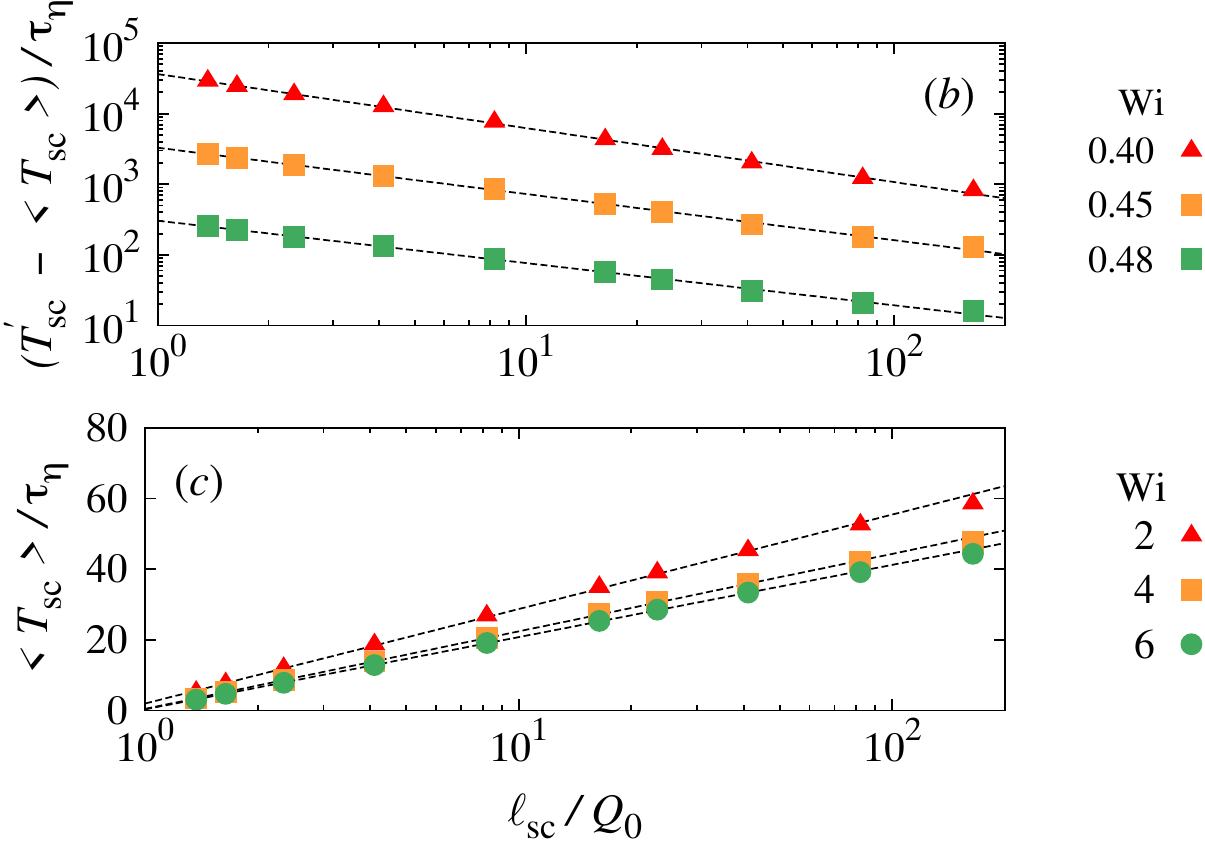}%
\caption{Passive polymers:
(\textit{a}) p.d.f. of the lifetime of a polymer for different values of
  $\Wi$. The inset compares the decay time of the fraction of unbroken polymers,
  the mean lifetime of a polymer, and the time scale $\gamma^{-1}$ in the exponential
  tail of the p.d.f. ($P(T_{\rm sc})\sim e^{-\gamma T_{\rm sc}}$ for $\gamma T_{\rm sc}\gg 1$) as a function of $\Wi/\Wi_{\rm cr}$;
  (\textit{b}) mean lifetime as a function of $\lbreak/Q_0$ below the coil-stretch transition. 
  $T_{\rm sc}'$ is a fitting parameter for the dashed curves.
  The data for $\Wi=0.4$ 
  (resp. $\Wi=0.45$) are multiplied by a factor of $10^2$ (resp.~10) in order to make the three curves more easily
  distinguishable;
  (\textit{c}) the same as in (\textit{b}) above the coil-stretch transition.}
\label{fig:lifetime}
\end{figure}

We now turn to the statistics of the lifetime $\Tbreak$ of a polymer.
The DNSs suggest that the
p.d.f. of $\Tbreak$ has an exponential tail
with a time scale $\gamma^{-1}$ that, beyond $\Wi_{\rm cr}$,
decreases rapidly as a function of $\Wi$ (figure~\ref{fig:lifetime}\textit{a}).
For all values of the Weissenberg number, 
$\gamma^{-1}$ is approximately the same as the
decay time of the fraction of unbroken polymers owing to
the exponential decay of the latter at long times (see \S~\ref{sect:analytical}).
However, $\gamma^{-1}$ differs
from $\langle T_{\rm sc}\rangle$, because the exponential behaviour of the
p.d.f. of $\Tbreak$ sets in only at relatively large values of $\Tbreak$.
For a fixed $\Wi$, the mean lifetime $\langle T_{\rm sc}\rangle$ behaves as a power
of $\lbreak/Q_0$ below the coil-stretch transition and as the
logarithm of $\lbreak/Q_0$ beyond that (see 
figures~\ref{fig:lifetime}\textit{b} and~\ref{fig:lifetime}\textit{c});
we remind the reader that $Q_0$ is the initial length of any link of the 
chain.
Small deviations 
are only observed for $\lbreak\gg Q_0$. 
Moreover, we have checked that, for $\Wi<\Wi_{\rm cr}$, the exponent $\beta$
that gives the dependence of $\langle\Tbreak\rangle$ on $\lbreak/Q_0$
is the same as the exponent that describes the right tail of $\widehat{P}(R)$,
i.e. $\widehat{P}(R)\propto R^{-1-\beta}$ for $r_0\ll R\ll(\mathscr{N}-1)
\lbreak$,
in agreement with \eqref{eq:tsc}.
Thus, the statistics of $\Tbreak$ in a turbulent flow is correctly described
by the predictions of \S~\ref{sect:analytical}.

We also note that the exponential tail of the distribution of $T_{\rm
sc}$ originates from the fact that scission is caused by the cumulative action
of the fluctuating strain-rate. This is in contrast to the 
fragmentation of sub-Kolmogorov inextensible fibres,
for which the internal tension depends on
the {\it instantaneous} velocity gradient projected along the fibre.
The p.d.f. of the scission time for fibres, therefore, reflects
the intermittent statistics of the velocity gradient and is strongly non-exponential
\citep*{sofia-2}.



\begin{figure}
\includegraphics[width=0.33\textwidth]{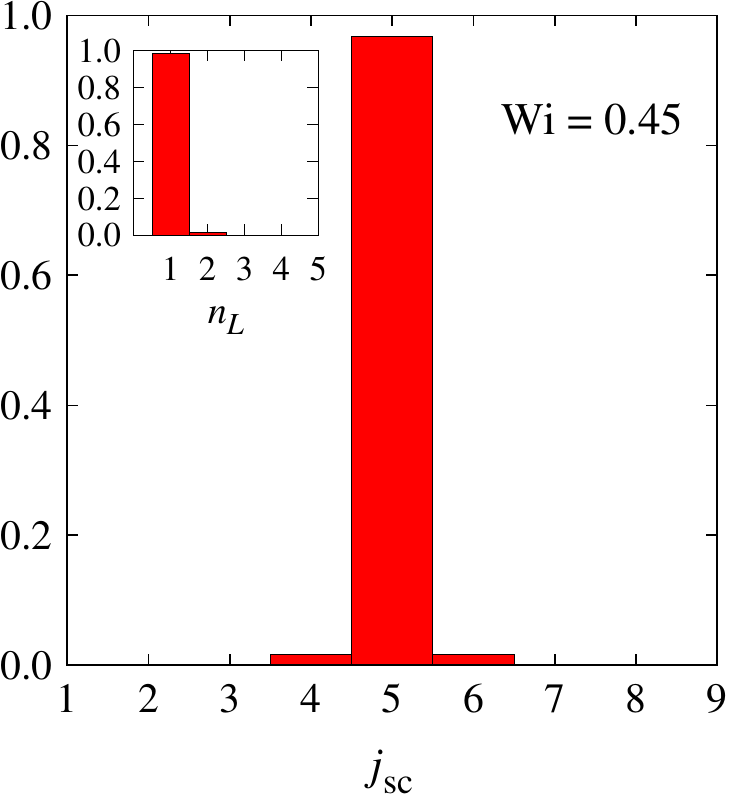}\hfill%
\includegraphics[width=0.33\textwidth]{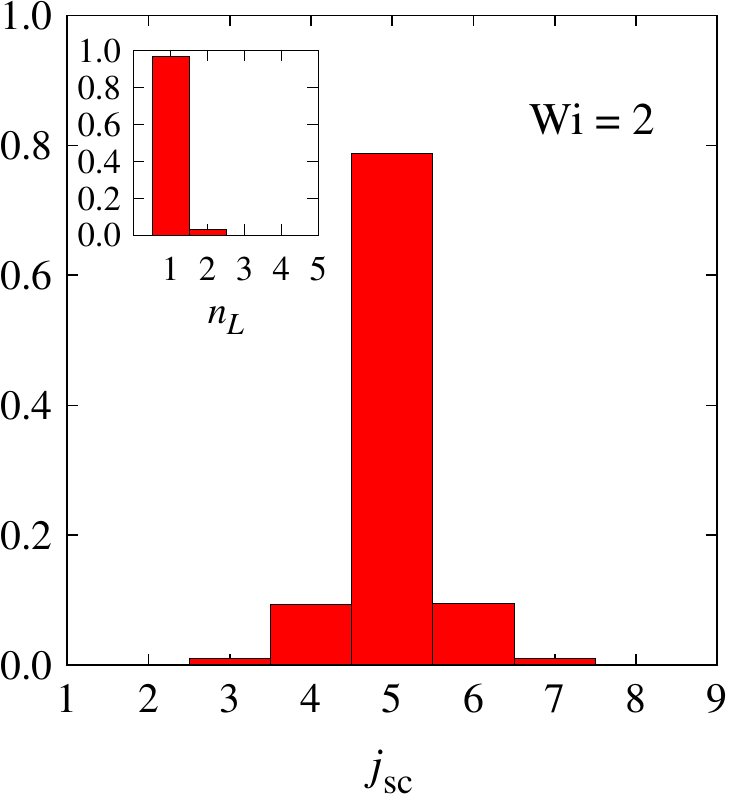}\hfill%
\includegraphics[width=0.33\textwidth]{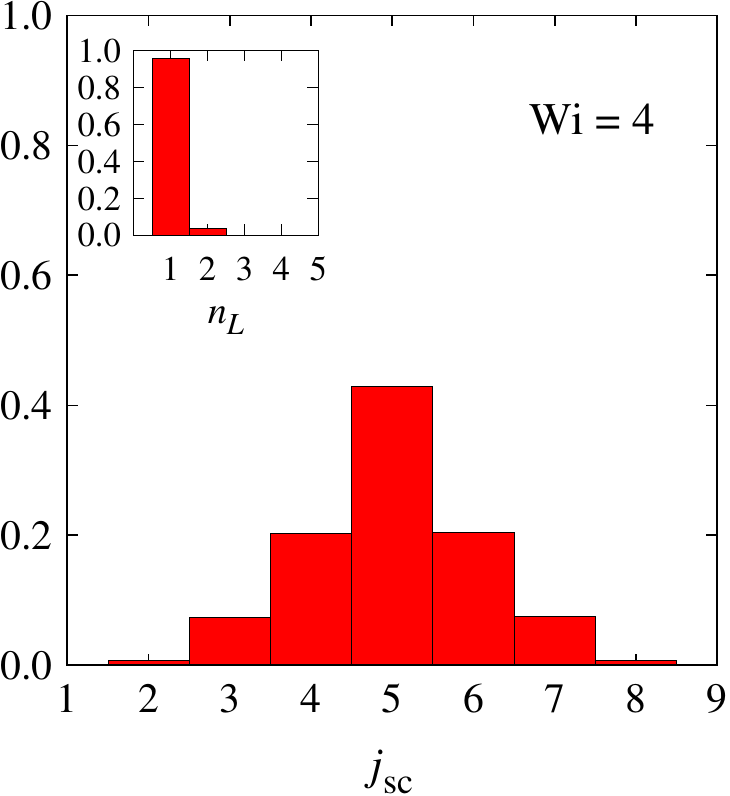}
\caption{Passive polymers:
probability of polymer scission occuring at the $j_{\rm sc}$-th link for different values of $\Wi$.
The insets shows the probability of $n_L$ links breaking simultaneously.}
\label{fig:histograms}
\end{figure}

Figure~\ref{fig:histograms} presents further results on
the statistics of the scission process.
As previously observed in experiments~\citep{hm84}, 
scission preferentially happens at
the midpoint of the polymer. However, the probability of scission happening
at the middle link decreases with $\Wi$. The reason for this is that, for small $\Wi$,
the chain is most of the time in a coiled state and scission occurs because of a sequence of
very strong fluctuations of $\bm\nabla\bm u$, whereas for large $\Wi$ all links
are consistently stretched near to the scission length.
The insets of figure~\ref{fig:histograms} show that
the probability of more than one link breaking simultaneously
is generally very small.

\begin{figure}
\centering
\includegraphics[width=0.485\textwidth]{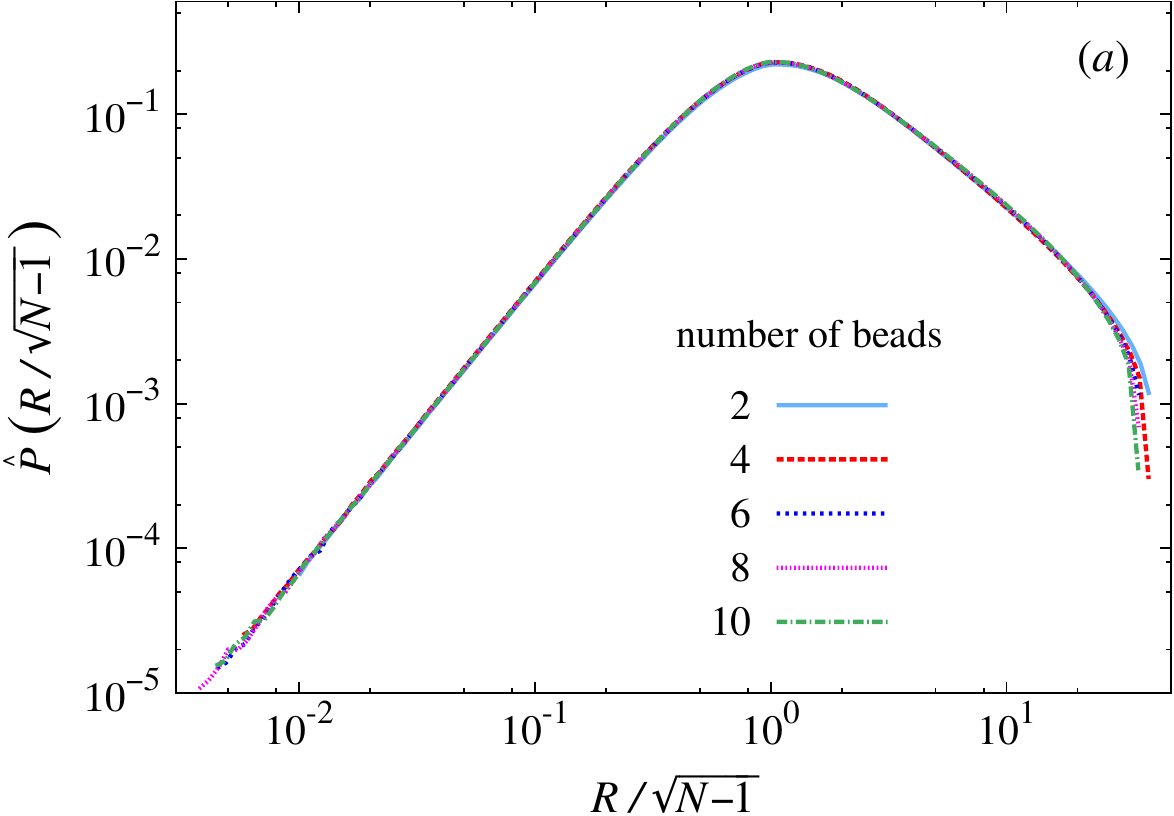}\hfill%
\includegraphics[width=0.5\textwidth]{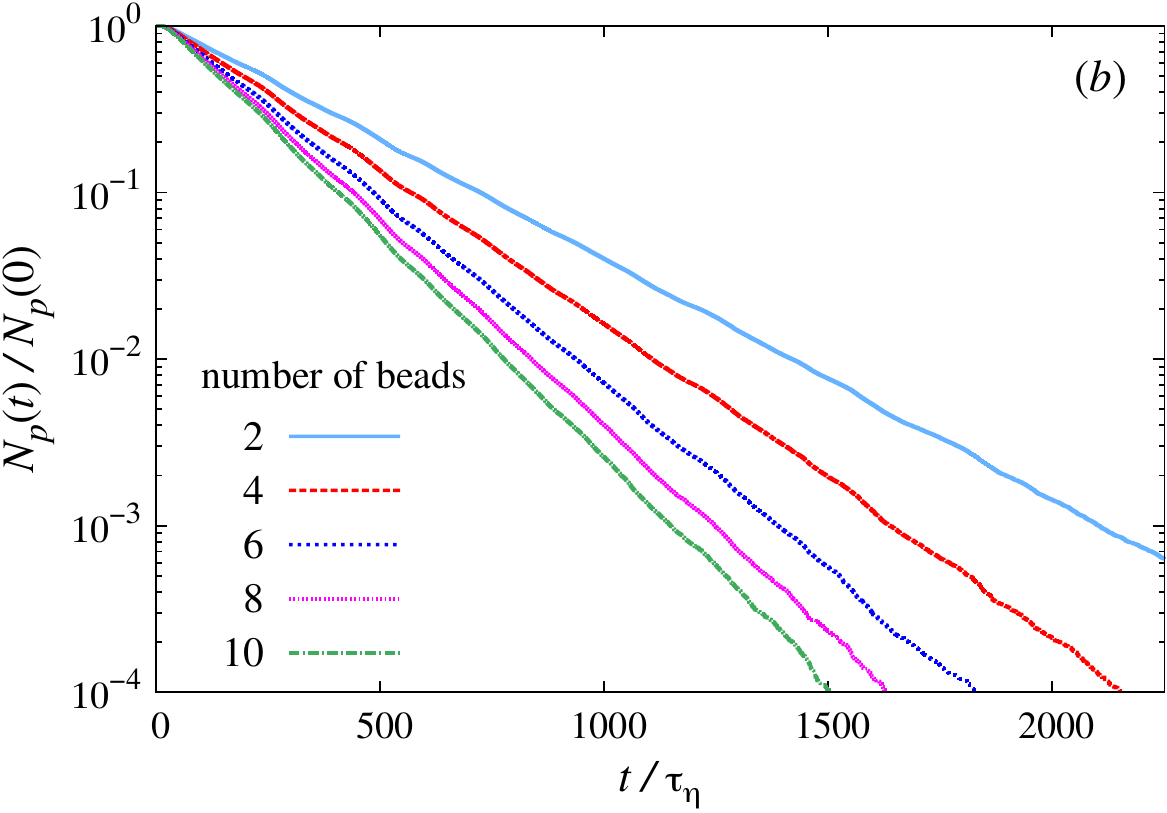}
\caption{Passive polymers:   (\textit{a})
time-integrated p.d.f. of the end-to-end separation of unbroken polymers
for $\Wi=0.6$ and different numbers of beads $\mathscr{N}$;
(\textit{b}) fraction of unbroken polymers as a function of time
for $\Wi=0.6$ and different numbers of beads $\mathscr{N}$.}
\label{fig:comparison}
\end{figure}

Finally, it was mentioned in \S~\ref{sect:analytical} that, in the absence of scission,
the dumbbell model ($\mathscr{N}=2$) captures the statistics of the
end-to-end separation of a full chain remarkably well,
provided the mapping in~\eqref{eq:mapping} is applied \citep[see][]{wg10}.
We have performed an analogous comparison in the presence of scission.
The time-integrated p.d.f.s 
in figure \ref{fig:comparison}\textit{a} show that, after the parameters of the chain
are suitably rescaled, the time-independent
statistics of the end-to-end distance is independent of
$\mathscr{N}$, except for small deviations close to the maximum extension.
Indeed, for a dumbbell, scission is defined in terms of the end-to-end
separation, whereas chains
with larger $\mathscr{N}$ break before all the links can
stretch up to $\lbreak$. 
These small differences have however a significant impact
on time-dependent quantities, such as 
the fraction of unbroken polymers:
small-$\mathscr{N}$ chains capture the temporal decay qualitatively,
but underestimate the scission rate
(see figure~\ref{fig:comparison}\textit{b}). 
The results also suggest that the discrepancies between
chains with $\mathscr{N}-1$ and $\mathscr{N}$ beads diminish
as $\mathscr{N}$ increases (figure~\ref{fig:comparison}\textit{b}) 
as well as when $\Wi$ increases (not shown).
We conclude that it is essential to consider the dynamics
of a full bead-spring chain in order to describe the scission
process correctly and achieve quantitative agreement between
experiments and constitutive models of polymer solutions.




\subsection{Active polymers}
\label{sect:active}

We now investigate the implications of the results obtained so far for the two-way-coupling
regime in which polymers perturb the surrounding flow.

When polymers break, their effective relaxation time $\tau^D$ decreases according to 
\eqref{eq:mapping} and the solution,
at any point in time, consists of polymers with different $\tau^D$.
We can then introduce a mean Weissenberg number $\langle\Wi\rangle (t)$,
which is defined as the average of $\lambda\tau^D$ over all polymers that
compose the solution at time $t$. 
Studying the evolution of $\langle\Wi\rangle(t)$, in a one-way-coupling simulation, helps us forsee how the effect of polymer-feedback on the flow would decay due to scission. 

\begin{figure}
  \centering
  \includegraphics[width=0.485\textwidth]{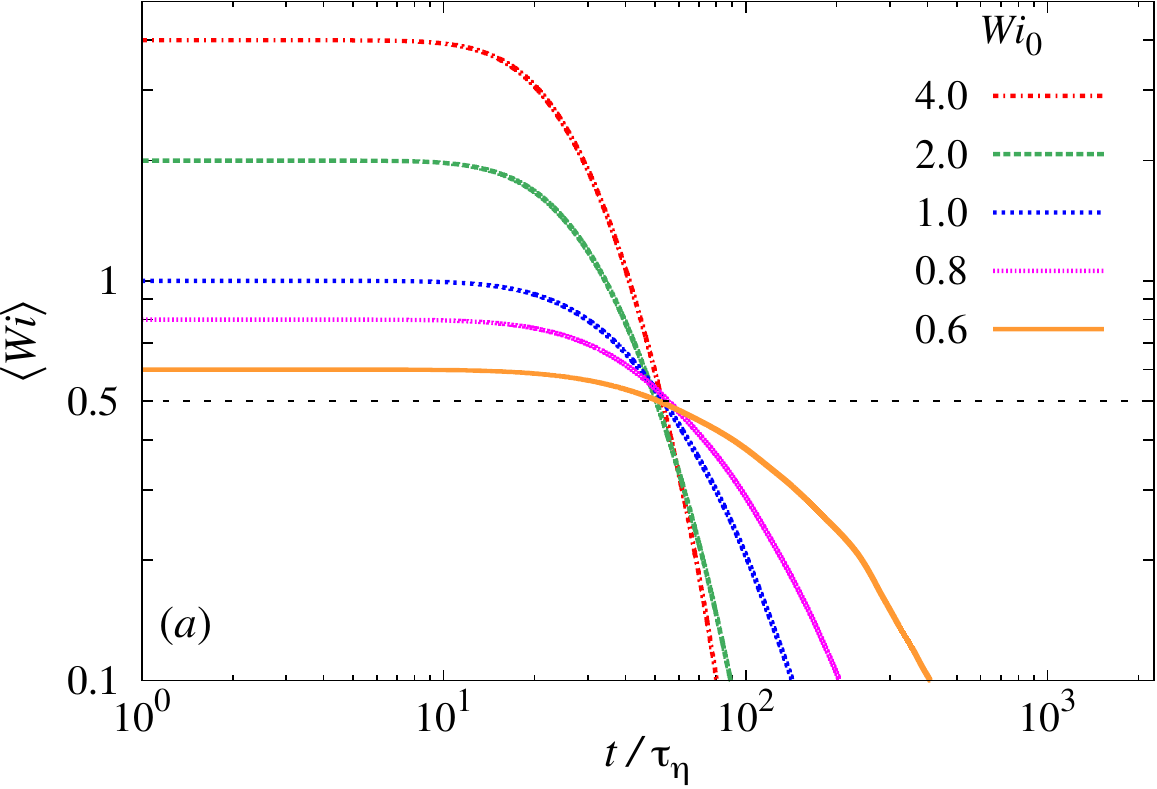}%
  \hfill%
  \includegraphics[width=0.5\textwidth]{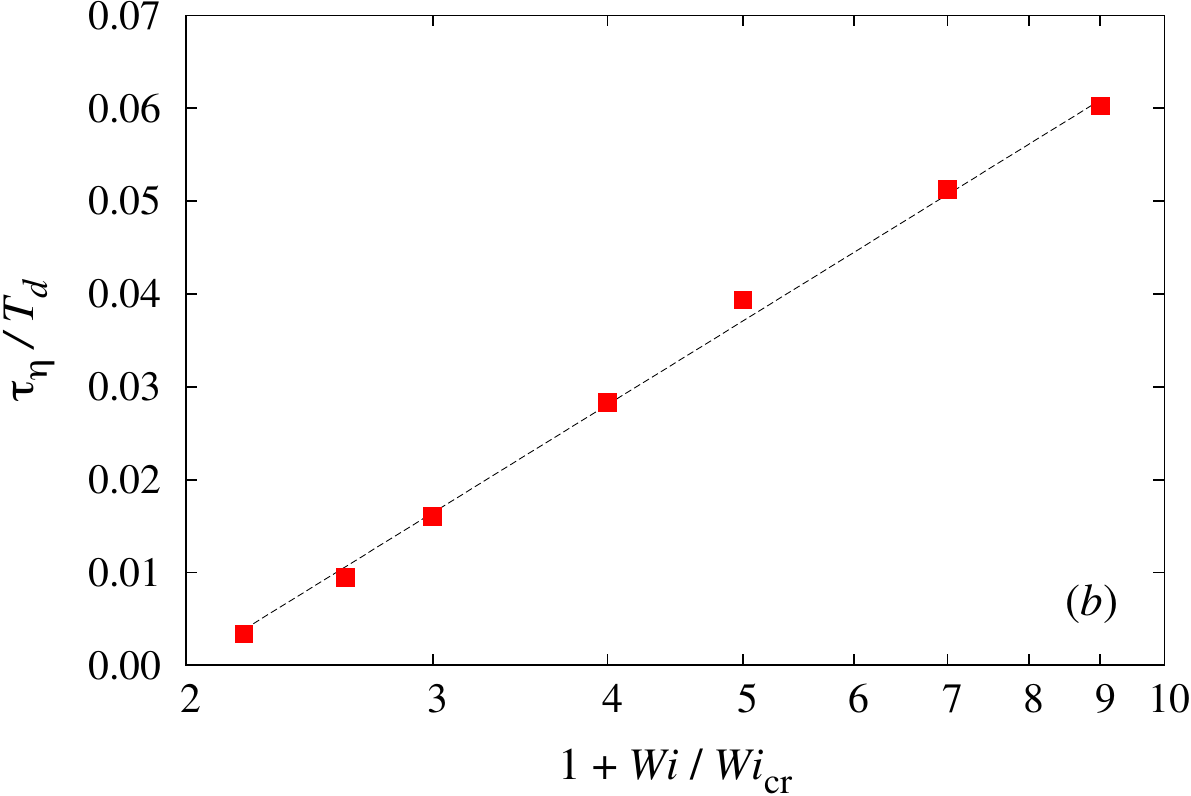}%
  \caption{Passive polymers: ({\it a}) mean Weissenberg number as a function of time
for $\mathscr{N}=2$ and different values of $\Wi_0$; ({\it b}) reciprocal of $T_d$
(multiplied by the Kolmogorov time scale) as a function of $1+\Wi/\Wic$.
The dashed line is
proportional to $\ln[( 1+ \Wi/\Wic)/2]$.}
  \label{fig:active-3}
\end{figure}

Let us, for the sake of simplicity, 
consider the case of dumbbells ($\mathscr{N}=2$) and take
an initial ensemble of $N_p(0)$ dumbbells with
Weissenberg number $\Wi_0$.
When a dumbbell breaks it forms two beads which formally have zero
Weissenberg number. Thus, at time $t$, the system consists of
$N_p(t)$ dumbbells with $\Wi=Wi_0$ and $2[N_p(0)-N_p(t)]$ single
beads with $\Wi=0$. Hence, for an ensemble of dumbbells,
\begin{equation}
\label{eq:mean-Wi}
\langle\Wi\rangle(t)= \dfrac{N_p(t)}{2N_p(0)-N_p(t)}\Wi_0.
\end{equation}
The temporal evolution of $\langle\Wi\rangle$ is obtained by calculating $N_p(t)$ from the Lagrangian
database used in \S~\ref{sect:sim}
and is shown in figure~\ref{fig:active-3}{\it a}
for different values of $Wi_0>\Wi_{\rm cr}$. Dumbbells with larger $\Wi_0$ have a larger
scission rate, and therefore $\langle\Wi\rangle$ vanishes rapidly.
In contrast, dumbbells with smaller $\Wi_0$ break relatively slowly and 
the mean $\Wi$ of the solution remains nonzero for a longer time.
We note, en passant, that $\langle\Wi\rangle$ becomes approximately equal to $\Wic$ 
at $t\approx 50\tau_\eta$ for all $\Wi_0$. By substituting \eqref{eq:N-decay} into
\eqref{eq:mean-Wi}, we thus deduce the following empirical expression for the
scission rate of dumbbells (see figure \ref{fig:active-3}{\it b}):
\begin{equation}
\frac{\tau_\eta}{T_d}\propto \ln\left[\frac{1}{2} \left( 1+ \frac{\Wi}{\Wic} \right)\right].
\end{equation}
The behaviour of $\langle\Wi\rangle$ shown in figure~\ref{fig:active-3}{\it a} 
suggests that a large value of $\Wi_0$ 
will produce a polymer feedback that is initially strong but short-lived, decaying rapidly due to scission. 
In contrast, a moderate value of $\Wi_0$
yields a feedback that, albeit weaker, should last for a longer time and may therefore be more effective.

To investigate this point in a two-way-coupling simulation,
we take the hybrid Eulerian--Lagrangian approach
proposed by \citet{wg13jfm,wg13b,wg14}, which consists
in seeding the fluid
with a large number of FENE dumbbells and calculating the reaction force
exterted by the dumbbells upon the fluid.
This amounts to adding the term $\bnabla\bcdot \mathsfbi{T}_p$ 
to the right-hand side of the Navier--Stokes
equations \eqref{eq:NS}, where $\mathsfbi{T}_p$ is 
the polymeric contribution to the stress tensor:
\begin{equation}
  \mathsfbi{T}_p=\dfrac{\nu\eta L^3}{N_p(0)}\sum_{n=1}^{N_p(0)}
  \dfrac{1}{\tau^{(n)}}
  \left[
    f\left(\dfrac{R^{(n)}}{Q_m^{D}}\right)
    \dfrac{3\bm R^{(n)}\otimes\bm R^{(n)}}{\req^2}
    -\mathsfbi{I}
    \right]
  \delta\left(\bm x-\bm X^{(n)}_c\right).
\end{equation}
In the expression of $\mathsfbi{T}_p$, $L$ is the linear size of the
domain and $\eta$ is the ratio of the polymer to the solvent contribution
to the total viscosity of the solution ($\eta$ is proportional to the 
volume fraction of dumbbells). The vectors $\bm X^{(n)}_c$ and
$\bm R^{(n)}$
are the positions of the centre of mass and the end-to-end
separation of the $n$-th dumbbell, respectively, 
$\mathsfbi{I}$ is the identity matrix, and
\begin{equation}
  \frac{1}{\tau^{(n)}}=
  \begin{cases}
    \tau^{-1} & \text{for $t < t^{(n)}_{\rm sc}$}
    \\
    0        & \text{for $t \geqslant t^{(n)}_{\rm sc}$,}
  \end{cases}
\end{equation}
where $t^{(n)}_{\rm sc}$ is the smallest time such that $R^{(n)}=\lbreak$.
Thus, the dumbbells stop affecting the velocity field after breaking.
The evolution of the position and the configuration of each dumbbell is
given by~\eqref{eq:dumbbell}.

The computational domain is a three-dimensional periodic box 
with $L=2\upi$. A pseudo-spectral method with $128^3$ grid points
and the second-order Runge--Kutta algorithm
are used for the integration of the Navier--Stokes equations
in space and in time, respectively.
The equation for the dumbbells are solved by using the Euler--Maruyama
scheme. The turbulent flow is maintained by a forcing with Fourier
transform $\widehat{\bm F}(\bm k,t)=A(t)\hat{\bm u}(\bm k,t)$
for $1\leqslant \vert\bm k\vert \leqslant 2$ and zero otherwise,
where $A(t)$ is such that the kinetic-energy input rate
$\epsilon_{\rm in}=\langle \bm u\bcdot\bm F\rangle$ is constant.
We take $\epsilon_{\rm in}=0.5$ and
$\nu = 0.015$, which yields $\mathit{Re}_\lambda
=51$ in the absence of polymer feedback
(for more details on the simulation, the reader is referred
to \citealt{wg13jfm}).

We first evolve the velocity field alone with $\eta=0$ and, once
a statistically steady flow is obtained, $N_p(0)=5\times 10^7$ 
polymers with $\eta=4\times 10^{-2}$ are dispersed
randomly into the fluid. In the following, the time $t=0$ corresponds to the
time at which polymers are added to the flow.
As in \S~\ref{sect:first-scission}, $Q^{D}_m/\req=\sqrt{3000}$,
$\lbreak=0.8Q_m^{D}$ and $r_0/\req=1$.
We shall consider
three values of the Weissenberg number $\Wi=\lambda\tau=0.6,0.8,1.0$,
where $\lambda$ is the Lyapunov exponent of the Newtonian flow ($\eta=0$).

\begin{figure}
  \centering
  \includegraphics[width=0.5\textwidth]{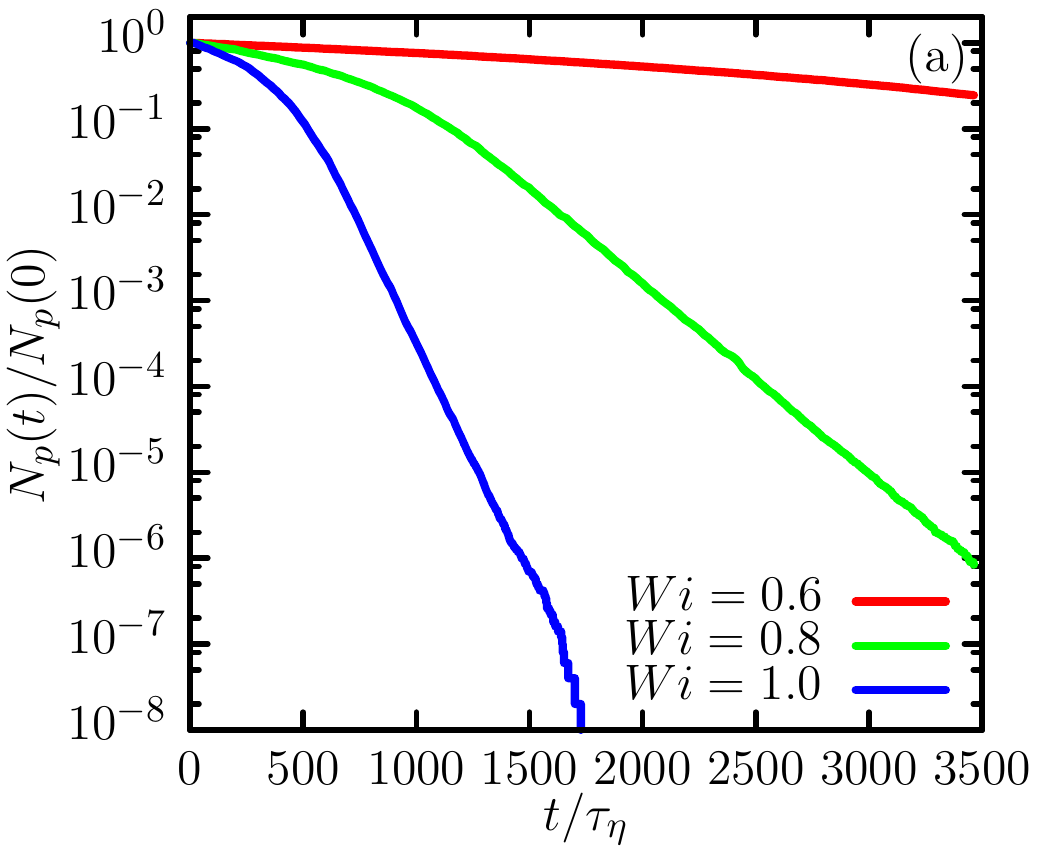}%
  \includegraphics[width=0.5\textwidth]{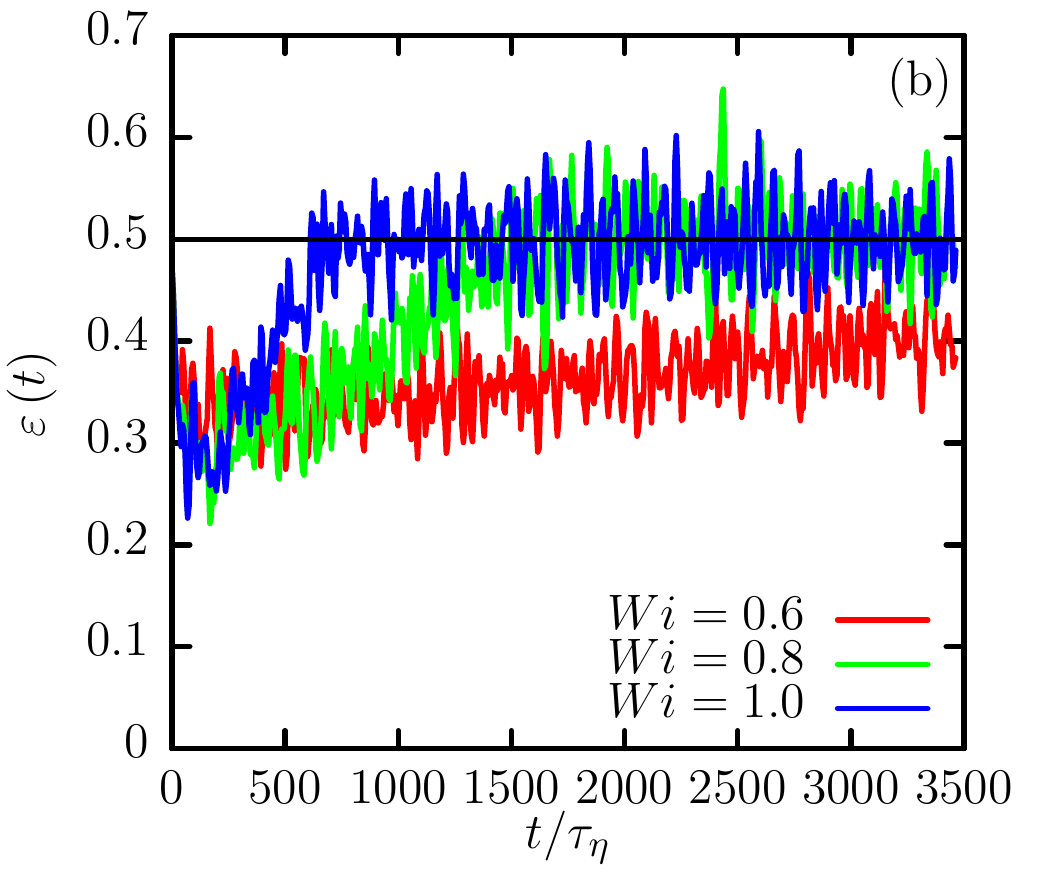}%
  \caption{Active polymers: ({\em a}) fraction of unbroken polymers
    as a function of time; 
    ({\em b}) time series of the kinetic-energy dissipation rate.
  }
  \label{fig:active-1}
\end{figure}

We begin by examining the time evolution of $N_p(t)/N_p(0)$.
Figure~\ref{fig:active-1}{\em a}
shows that initially the fraction of unbroken polymers
decreases slowly. However, at longer times and for large enough $\Wi$,
the decay becomes exponential.
The time at which the exponential decay sets in decreases rapidly with $\Wi$, while the corresponding decay rate increases.
This behaviour is the consequence of the way
the feedback of an ensemble of polymers evolves as a result of scission.
In isotropic turbulence, the dispersion of polymers into a Newtonian
solvent reduces the kinetic-energy dissipation rate by a factor proportional
to the polymer concentration \citep*{chirag,de_angelis,pmr06,pmr10,wg13jfm}.
The time series of the
instantaneous kinetic-energy dissipation rate
\begin{equation}
\epsilon(t)=\frac{\nu}{L^3}\int_{[0,L]^3} \vert\nabla\bm u(\bm x,t)\vert^2 
\mathrm{d}\bm x
\end{equation}
is shown in figure~\ref{fig:active-1}{\em b}
for different values of $\Wi$.
At short times, the polymer feedback is fairly strong and
causes a significant reduction
of $\epsilon(t)$ compared to the value of the Newtonian flow.
The reduction of $\epsilon(t)$ is associated with a corresponding
reduction of the 
amplitude of the velocity gradient and hence entails a weaker polymer
stretching and a lower probability of scission. For this reason,
the scission rate is initially small.
As time progresses, however, the concentration of unbroken polymers
keeps decreasing, until the effect of polymers becomes negligible
and $\epsilon(t)$ returns to its Newtonian value.
This is accompanied by a growth of the velocity gradient and thus a faster
decay of the fraction of unbroken polymers.
Since for higher values of $\Wi$ polymers break more easily,
the time needed for the polymer feedback to vanish decreases
with increasing $\Wi$. So for $\Wi=1$ the reduction of energy
dissipation is initially stronger, but the return to the
Newtonian regime is fast; for $\Wi=0.6$ the same process is
extremely slow and the dissipation reduction, albeit smaller,
lasts for a much longer time.

One important consequence of this analysis is that, for a given experiment, an optimal, not necessarily very large, $\Wi$ 
should maximize the energy-dissipation reduction integrated over the duration of the experiment.

\section{Multiple scissions}

We now return to passive polymers and examine how a population evolves when polymers can undergo multiple, repeated
scissions. 
When a polymer composed of several beads undergoes its first scission (whose statistics was analysed in \S~\ref{sect:first-scission}), it results in two daughter polymers, each containing a smaller number of beads. These daughter polymers can themselves undergo further scissions to produce a tertiary generation, and so on, until, in case of complete breakage, we are left with only individual beads which represent small, inextensible polymer fragments. In this section, we study how this hierarchy of daughter polymers evolves, owing to multiple breakups of the parent polymers. After every scission event, we discard the broken link, form two new daughter polymers, and then follow their evolution along the trajectories of their respective centres of mass, which typically separate exponentially in time due to the positive Lyapunov exponent of the turbulent flow.

In the simulations for first-scission statistics, described earlier in \S~\ref{sect:sim}, we treated the centre of mass and the separation vectors of the polymer chains as the dynamical variables. This approach becomes inconvenient when dealing with multiple scissions, because new trajectories would have to be spawned after each scission event. So, we instead adopt the following approach. Consider $N_p(0)$ parent polymer chains, each of which is composed of $\mathscr{N}_0$ beads. Thus, the total number of beads in the simulation, which remains constant in time, is $N_b=\mathscr{N}_0 N_p(0)$. We assign to the beads distinct labels from 1 to $N_b$ and follow the time evolution of their positions. In addition, we maintain an array which records the labels of the first and last beads of every polymer chain in the simulation. So when a polymer chain undergoes a scission, we simply update this array: the broken chain is assigned all the beads preceding the broken link (and thereby becomes the first daughter polymer), while the remaining beads are assigned to a new entry in the array (to form the second daughter polymer). This procedure allows us to simulate the growing population of polymer chains without increasing the number of dynamical variables. For a given chain with $\mathscr{N}$ beads,
we still use \eqref{eq:chain} to evolve its dynamics, 
by calculating $\bm X_c$ and $\bm Q_i$ from the position vectors of the beads, $\bm X_1,\dots,\bm X_{\mathscr{N}}$:
\begin{equation}
{\bm X}_c = \sum_{i=1}^{\mathscr{N}}\bm X_i \quad {\rm and} \quad {\bm Q}_i = {\bm X}_{i+1} - {\bm X}_{i} \;\; (1 \le i \le \mathscr{N}-1).
\end{equation} 
All other aspects of these simulations follow the description given for the first-scission simulations in \S~\ref{sect:sim}, except that we now simulate the turbulent flow with a more moderate value of $Re_{\lambda} = 90$, by using $256^3$ collocation points and a time step of ${\rm d}t = 10^{-3}$. The reason for this is that multiple scission simulations require the polymers to be evolved simultaneously along with the flow (whereas a precomputed set of tracer trajectories were used to obtain all first-scission statistics), thus necessitating a new flow computation for each variation of the polymer parameters.

\begin{figure}
\begin{center}
\includegraphics[width=0.95\textwidth]{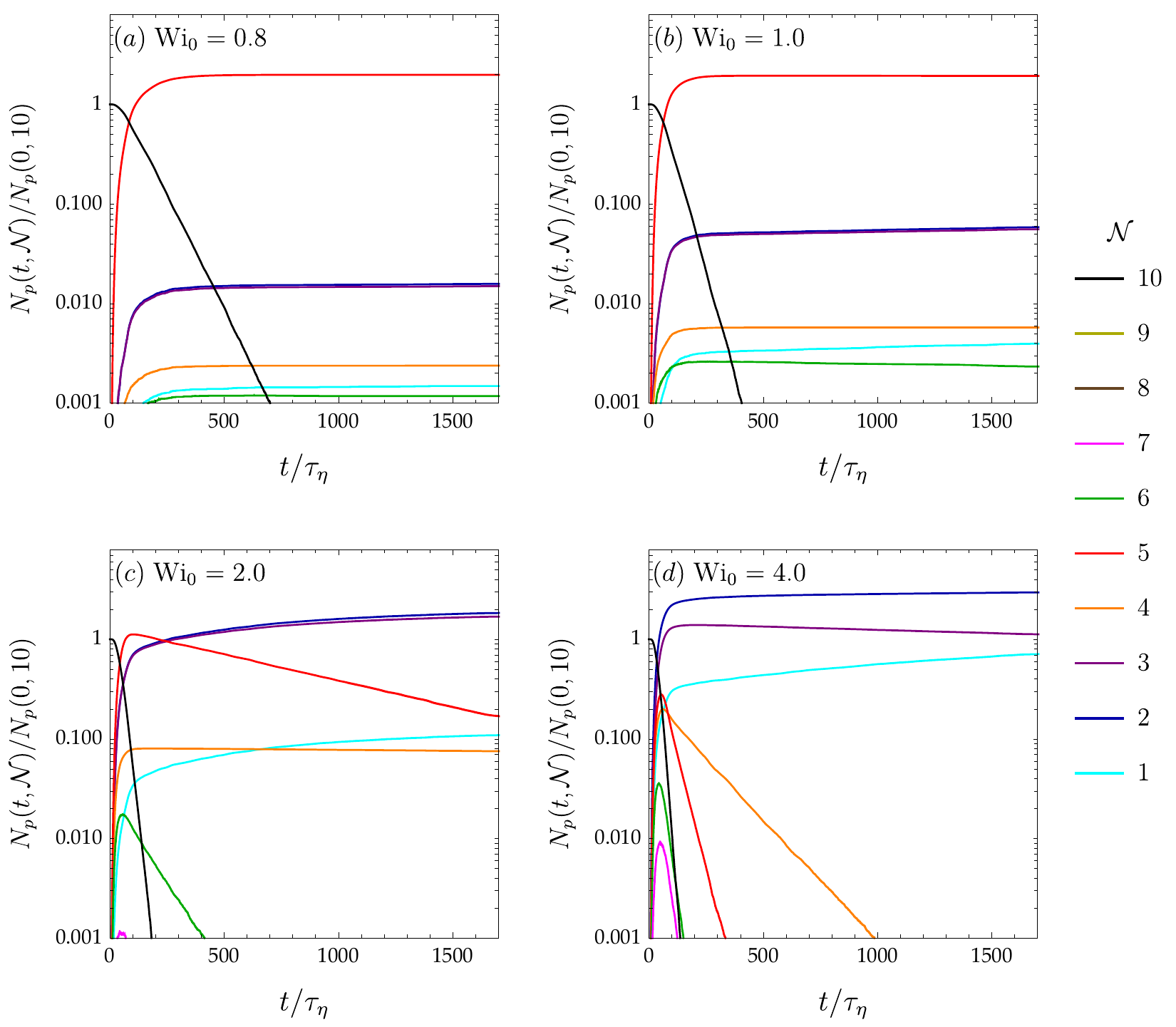}
\end{center}
\caption{Passive polymers undergoing multiple scissions: evolution of the number of polymers, $N_P (t,\mathscr{N})$, categorized according to the number of constituent beads $\mathscr{N}$, due to the repeated scission of $N_P (0,10)$ 10-bead parent polymers with (\textit{a}) $\Wi_0=0.8$, (\textit{b}) $\Wi_0=1.0$, (\textit{c}) $\Wi_0=2.0$ and  (\textit{d}) $\Wi_0=4.0$.}
\label{fig:decay_mult}
\end{figure}

At time $t=0$, the parents polymers are composed of $\mathscr{N}_0$ beads. 
As a result of scissions, at later times polymers with different numbers of beads will be found in the flow.
We thus denote by $N_P(t,\mathscr{N})$ the number of polymers that, at time $t$, are composed of $\mathscr{N}$ beads.

We begin by examining how $N_P(t,\mathscr{N})$ evolves due to the repeated scission of an initial population of $N_P(0,\mathscr{N}_0)$ parent polymer chains, each with $\mathscr{N}_0=10$ beads.  Figure~\ref{fig:decay_mult} presents results for various values of the parent-polymer Weissenberg number,  $\Wi_0$ = (\textit{a}) 0.8, (\textit{b}) 1.0, (\textit{c}) 2.0, and (\textit{d}) 4.0. Each $\mathscr{N}$-bead parent/daughter polymer is represented by a different curve, as indicated in the legend. We observe that for the small values of $\Wi_0$ in panels (\textit{a}) and (\textit{b}), only 10-bead polymers break. 
As shown in figure~\ref{fig:histograms}, most of the daughter polymers have 5 beads (along with a few non-5-bead polymers due to rare off-centre scissions),
because scission typically occurs at the central link for small $Wi_0$.
For the larger values of $\Wi_0$ in panels (\textit{c}) and (\textit{d}), the daughter polymers also undergo scissions and the population is eventually dominated by small polymers and individual beads (inextensible polymer fragments).

Clearly, the extent to which polymers can undergo repeated scissions depends on the $\Wi_0$ of the parent polymers. This is because each subsequent scission produces daughter polymers with a smaller relaxation time. This can be quantified by using \eqref{eq:mapping} to calculate the effective Weissenberg number of the $\mathscr{N}$-bead daughter polymer:
\begin{equation}
\Wi = \frac{\mathscr{N} (\mathscr{N}+1)}{\mathscr{N}_0 (\mathscr{N}_0+1)} \Wi_0,
\label{eq:WiN}
\end{equation}
where $2 \leq \mathscr{N} < \mathscr{N}_0$, and $\Wi_0$ is the Weissenberg number of the parent polymer with $\mathscr{N}_0$ beads. Note that $\Wi \equiv 0$ for individual beads ($\mathscr{N} = 1$). An approximate condition for a significant fraction of any given generation of daughter polymers to breakup is $Wi > Wi_{c} = 1/2$. So, if $\mathscr{N}_0 = 10$, \eqref{eq:WiN} implies that 5-bead daughter chains will not breakup, unless $\Wi_0 > 1.83$. The results in figure~\ref{fig:decay_mult} agree with this estimate: 5-bead daughters experience a slow rate of scission for $\Wi_0 = 2.0$ (panel ({\em c})), but no scission at all for $\Wi_0$ = 0.8 and 1.0 (panels ({\em a}) and ({\em b}) respectively). Furthermore, for $\Wi_0 = 4.0$, we expect scissions of daughter polymers that have $\mathscr{N} = 4$ beads or more, but certainly not if they are composed of only 2 beads. This is exactly what we observe in figure~\ref{fig:decay_mult} (panel ({\em d})). Thus, the condition $\Wi > Wi_{\rm cr}$, in conjunction with \eqref{eq:WiN}, provides a simple way of estimating the number of beads of the smallest polymer that can be formed by repeated scission, given $\mathit{Wi}_0$.

Panels ({\em c}) and ({\em d}) of figure~\ref{fig:decay_mult} show that the evolution of the number of daughter polymers typically have two regimes. First, there is an enrichment phase, during which daughter polymers are formed due to the rapid breakup of the 10-bead parent polymer. Daughter polymers with various number of beads can be formed, especially at large $Wi$ for which off-centre breakups are quite frequent (figure~\ref{fig:histograms}). An exponential decay phase then appears for the daughter polymers that have enough beads to undergo further scission, e.g., $\mathscr{N} = 7, 6, 5, 4$ in panel ({\em d}). These secondary breakups produce a second phase of mild enrichment for the smallest polymers that do not breakup, e.g., $\mathscr{N} = 2$ in panels ({\em c}) and ({\em d}). 

\begin{figure}
\begin{center}
\includegraphics[width=\textwidth]{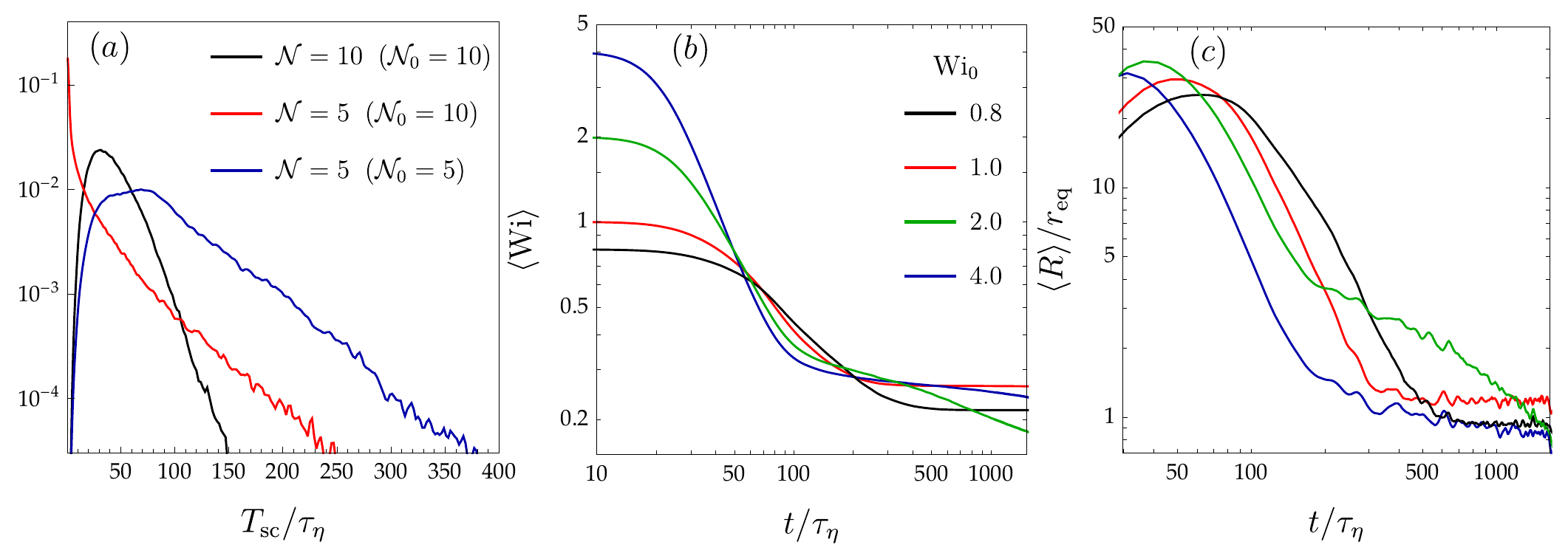}
\end{center}
\caption{Passive polymers undergoing multiple scissions: (\textit{a}) probability distributions of the lifetime of a parent 10-bead polymer, a daughter 5-bead polymer formed because of the scission of a 10-bead polymer ($\mathscr{N}_0 = 10$) and a parent 5-bead polymer ($\mathscr{N}_0 = 5$) with same $\Wi_0=4$; (\textit{b}) decay of $\Wi$, averaged over the entire population of polymers, for various values of the initial $\Wi_0$ of the 10-bead parent polymers; (\textit{c}) evolution of the averaged end-to-end extension $\langle R \rangle$ of the polymers for various $\Wi_0$ of the parent polymers. }
\label{fig:Tsc_mult}
\end{figure}

The decay time $T_d$ of the number of daughter polymers, in the exponential decay regime, is larger for smaller $\mathscr{N}$, as evidenced by the shallower slopes of the corresponding curves (e.g., $\mathscr{N} = 7, 6, 5, 4$ in panel ({\em d})). This variation follows from figure~\ref{fig:decay}\textit{b}, provided that the decay time of each daughter polymer can be estimated from figure~\ref{fig:decay}\textit{b} by
using \eqref{eq:WiN} to calculate $\Wi$. However, the use of figure~\ref{fig:decay}\textit{b} for daughter polymers is permissible only if the scission kinetics of a $\mathscr{N}$-bead daughter polymer is the same as the first-scission kinetics of a parent polymer that starts with $\mathscr{N}$ beads. To check if this is the case, we ran a separate simulation to calculate the distribution of survival times of 5-bead parent polymers ($\mathscr{N}_0 = 5$, $\mathscr{N} = 5$). In figure~\ref{fig:Tsc_mult}\textit{a}, these results are compared with the survival time distribution of 5-bead daughters, which form from the breakup of 10-bead parent polymers. The corresponding distribution for the 10-bead parent polymers is also shown for comparison. Remarkably, we see that daughter polymers have a much higher probability of breaking at early times than parent polymers with the same number of beads. This is because daughter polymers are formed from scission events:
 they typically start out in a stretched configuration and so have a much higher probability of breaking quickly than a randomly initialized, parent polymer with the same number of beads. The effect of this fades quite quickly, though, and for times larger than the Lagrangian decorrelation time $T_L \approx 10 \tau_\eta$ 
(for $Re_{\lambda} \sim 10^2$; see \citealt{Yeung2002}) the survival time p.d.f of the daughter polymers begins to resemble that of the parent polymers with the same number of beads. The time scales associated with the two exponential tails is therefore approximately the same, and one can indeed estimate the long-term decay time scale (or equivalently the decay rate) of an $\mathscr{N}$-bead daughter polymer, formed due to multiple scissions, by using just the first-scission statistics of a parent polymer with the same number of beads.

Can the early-time, high scission probability of daughter polymers be ignored? This depends on the $\Wi$ of the daughter polymers. From figure~\ref{fig:decay}\textit{b}, we expect $T_d \gg T_L$ for small $\Wi \lesssim 1$; the number of polymers that would breakup during the initial time $T_L$ will then be relatively small, and may be ignored. However, for larger 
$\Wi \gtrsim 4$, we have $T_d \approx T_L$ and a significant fraction of daughter polymers would break before they forget their stretched initial conditions. Therefore, for large $\Wi_0$ (which would produce daughter polymers with large $\Wi$), it becomes important to simulate multiple scissions in order to faithfully describe the decay of the polymer population. Such simulations would have to be repeated if the number of beads of the parent polymer changes. The situation simplifies considerably, however, for small $\Wi_0$ as one can then use first-scission statistics, calculated for a range of $\mathscr{N}$, to describe the decay of daughter polymers, regardless of the number of beads of the parent polymer.

The evolution of $\Wi$ averaged over all polymers, which represents the mean effective $\Wi$ of the entire population, is shown in figure~\ref{fig:Tsc_mult}\textit{b}, for various values of $\Wi_0$ of the parent polymers. This is the analogue of figure~\ref{fig:active-3}, 
but for $\mathscr{N}_0 = 10$ rather than $2$. We see that $\langle \Wi \rangle$ decreases rapidly while the parent polymers are breaking. After this, the slower 
scission rates of the daughter polymers lead to a more gradual decrease in $\langle \Wi \rangle$, which eventually will saturate to a value less than $Wi_{\rm cr}$, corresponding to daughter polymers that cannot be broken up further by the flow. 
This effect cannot be captured by dumbbells, which directly breaks into
beads (figure~\ref{fig:active-3}). 

The reduction of $\langle \Wi \rangle$ with time has a strong influence on the evolution of the mean end-to-end extension of the polymers $\langle R \rangle$, as shown in figure~\ref{fig:Tsc_mult}\textit{c}. After an initial stretching phase, $\langle R \rangle$ begins to decrease, owing to both the scission of highly 
stretched polymers and the smaller $\Wi$ of the resulting daughter polymers. After all scissions cease, we are left with relatively in-extensible daughter polymers whose extensions fluctuate near $r_{eq}$. Interestingly, the curve for $\Wi_0 = 2.0$ shows two regimes in the decay of $\langle R \rangle$. The first fast-decay regime is due to the rapid scission of 10-bead parents (see figure~\ref{fig:decay_mult}\textit{c}); the second slow-decay regime results from the much slower breakup of 5-bead daughter polymers, for which $\Wi = 0.55$ only {\it just} 
satisfies the condition for secondary scission ($\Wi > Wi_{\rm cr} = 1/2$). This second slow regime is not seen for either smaller or larger $Wi_0$: in the former case, secondary breakups do not occur, whereas in the latter case secondary breakups occur very quickly (see figure~\ref{fig:decay_mult}).

\section{Concluding remarks}

Polymers, even in small quantities, have a dramatic impact on a turbulent flow, reducing drag or dissipation and suppressing small-scale motion. However, because these effects
originate from polymer stretching (and the resultant feedback forces), the polymers which exert the strongest influence on the flow are also the most susceptible to strain-induced
scission. Therefore, to achieve effective flow modification, one must strike a balance between these opposing tendencies, which in turn demands a detailed understanding of the scission
process and the factors that influence the rate of scission.

In this work, we have analysed the scission of polymers in homogeneous isotropic turbulence, with a focus on the temporal decay of unbroken polymers, and the statistics of their survival
times. By using direct numerical simulations, we have quantified the decay time (or scission rate) as a function of $Wi$, which can serve as inputs for coarse-grained models. Importantly,
all the key qualitative features of the numerical results can be predicted analytically by replacing the fluctuating, turbulent velocity gradient by a time-decorrelated Gaussian random
flow. 
This is possible
because scission is caused by the cumulative action of fluctuating strain, and \textit{not} by sudden stretching in high-strain regions of the flow. 
Another finding, relevant for future
computations, is that a multi-bead polymer chain cannot be replaced by a dumbbell model without incurring quantitative
errors in the prediction of breakup rates. However, the results appear to
converge as $N_b$ increases, so we expect $N_b \sim O(10)$ to be sufficient even if a polymer model may strictly demand many more beads.

Our study of the scission of active polymers has shown that there is an intermediate value of $\Wi$ for which the overall, time-integrated, reduction
of the kinetic-energy dissipation rate is maximum: for small $\Wi$
the polymers do not stretch and the feedback is weak, whereas for large $Wi$ the stretching and feedback is
initially strong, but the resultant dissipation-reduction is lost rapidly as the polymers breakup very quickly.

We have shown that in a sufficiently strong turbulent flow (large $Wi$) polymers can breakup repeatedly. However, because the fragments in each successive generation have a smaller relaxation time, the breakup process eventually ceases once the effective $Wi$ of the surviving polymer fragments becomes less that $Wi_{cr}$. From this condition, we can estimate the number of beads in the largest surviving chains as $\mathscr{N} (\mathscr{N}+1) \le 6 Wi_{\rm cr} \tau_{\eta}/\tau$ [using ~\eqref{eq:mapping}]. Now, as $\mathscr{N}$ is linearly related to the mass of a polymer, this condition allows us to estimate how the weight-averaged molar mass (biased towards the largest chains) of the surviving polymers $M_{\rm ws}$ scales with the Reynolds number $Re$. For large $\mathscr{N}$, we have $\mathscr{N}^2 \sim \tau_{\eta}/\tau \sim Re^{-3/2} \nu/D^2 \tau$, where $D$ is the large length-scale of the flow system. Therefore, for a specified polymer, solvent and system geometry we obtain $M_{\rm ws}^2 \sim Re^{-3/2}$. This scaling is consistent with the experimental data of \cite{vcs06}, who obtain a power-law exponent close to $-3/2$ for a variety of polymers and system geometries (these results are reported in terms of the squared, weight-averaged polymer length which is linearly related to $M_{\rm ws}^2$, as described in the Data Analysis section of \citealt{vcs06}).

The multiple-scission statistics of a polymer also show a non-monotonic dependence on $\Wi$. Small $\Wi$ polymers break only once, if at all, whereas large $\Wi$ polymers undergo a
rapid succession of breakups and quickly reach their limiting generation (fragments which are no longer stretched by the flow). However, for intermediate $\Wi$, the first-scission
occurs quickly, but then the daughter polymers breakup much more slowly. This introduces multiple time-scales into the decay of $\langle Wi \rangle$, the average effective $Wi$ of the
polymer population. This average quantity and its evolution are relevant to coarse-grained continuum models of polymer solutions which typically contain a single mean
polymer-relaxation-time parameter. Indeed, the development of continuum models that incorporate scission is essential for predicting the long-time dynamics of turbulent polymer
solutions in complex applications. The quantitative results as well as physical insights gained from this study should aid in the development of such models.

\begin{acknowledgments}
  JRP, SSR 
  and DV thank the Indo--French Centre for Applied Mathematics (IFCAM)
  for financial support. 
DV acknowledges his Associateship with the International Centre for Theoretical Sciences, Tata Institute of Fundamental Research, Bangalore, India.
  The computations were perfomed at Centre  de  Calculs  Interactifs of
  Universit\'e  C\^ote d'Azur, at
Information Technology Centre in Nagoya University and National Institute for Fusion Science (NIFS),
and at the cluster Mowgli and the work station Goopy at the ICTS--TIFR.
  TW thanks the financial support by MEXT KAKENHI through Grant No.~20H00225 and JSPS KAKENHI Grants No.~18K03925, and the computational supports provided by JHPCN (jh190018-NAH, jh200006) and by NIFS (NIFS18KNXN366, NIFS18KNSS105).
SSR acknowledges support of the DAE,
Govt. of India, under project no.~12-R\&D-TFR-5.10-1100 and  DST (India) project MTR/2019/001553 for support.
  JRP acknowledges funding from the IIT Bombay IRCC Seed
  grant.
\end{acknowledgments}


%


\end{document}